\documentclass[aps,prb,amsmath,amssymb,reprint,superscriptaddress,footinbib]{revtex4-2}

\usepackage{graphicx}
\usepackage{float}
\usepackage{bm}
\usepackage[colorlinks,allcolors=blue]{hyperref}
\usepackage{siunitx}
\usepackage{natbib}
\usepackage{bold-extra}
\usepackage[normalem]{ulem}

\begin{document}
	
\title{Phase-dependent supercurrent and microwave dissipation of HgTe quantum well Josephson junctions}

\date{\today}

\author{Wei Liu}
%\thanks{These two authors contributed equally}
\email{wei.liu@uni-wuerzburg.de}
\author{Stanislau U. Piatrusha}
%\thanks{These two authors contributed equally}
\email{stanislau.piatrusha@uni-wuerzburg.de}
\author{Lena Fürst}
\author{Lukas Lunczer}
\author{Tatiana Borzenko}
\author{Martin P. Stehno}
\author{Laurens W. Molenkamp}
\affiliation{Experimentelle Physik III, Physikalisches Institut, Universit\"{a}t W\"{u}rzburg, Am Hubland, 97074 W\"{u}rzburg, Germany.}
\affiliation{Institute for Topological Insulators, Universit\"{a}t W\"{u}rzburg, Am Hubland, 97074 W\"{u}rzburg, Germany}

\begin{abstract}
We measured the microwave response of a HgTe quantum well Josephson junction embedded into an RF SQUID loop which is inductively coupled to a superconducting resonator. The side-contacted devices studied here operate in bulk transport mode, with a separation between the superconducting contacts smaller than both the estimated carrier mean free path and superconducting coherence length. We extract the current-phase relation and the phase-dependent microwave dissipation, that, at low temperature, primarily is related to photon-induced transitions between the Andreev bound states. We study the effects of gate voltage and temperature on our devices and compare the measurements with a tight-binding model based on the Bogoliubov-de Gennes equations. A combined analysis of both microwave admittance components allows us to confirm the presence of a small gap in the Andreev bound state spectrum at phase $\pi$, indicating high interface transparency, matching our observations in DC measurements of a similar device. Our work demonstrates the versatility of microwave measurements as a tool for Josephson junction characterization and highlights the importance of the interface properties for side-contacted Josephson devices.
\end{abstract}

\maketitle

\section{Introduction}
Hybrid Josephson junctions (JJs) with topological materials as a weak link provide a promising platform for the realization of exotic superconducting states~\cite{FuKane2008, FuKane2009, Frolov2020}. These arise from topological properties of the band structure~\cite{Bernevig2006, Konig2007} in conjunction with the possibility of manipulating the electrons in the semiconductor material by electric and magnetic fields. A quantum-mechanical description of the JJ can be given in terms of Andreev bound states (ABS). The spectrum of the ABS and their occupation determine the supercurrent flow~\cite{kulik1969macroscopic}. Known topological materials are very sensitive to growth and device processing conditions, thus Josephson devices may not be in the topological regime (i.e., in the band gap of the material) but feature a parallel transport channel of trivial ABS in the bulk~\cite{Hart2014, Bocquillon2017}. Here, we characterize JJs with weak links made from HgTe quantum wells (QWs). We use wet etching to minimize sample damage and push for a sub-500\,$\unit{\nano\meter}$ superconducting electrode separation, aiming for the short junction limit.

In low-frequency circuits, it often suffices to know the current-phase relation (CPR)\,\cite{Golubov2004} that relates the supercurrent to the superconducting phase difference across the JJ for a complete description of the relevant properties. The CPR is directly related to the ABS spectrum and may carry over topological properties of the ABS~\cite{Peng2016}. A measurement of the CPR itself, however, requires fixing the phase difference across the junction. This can be accomplished by incorporating the JJ with a second junction in an asymmetric DC SQUID and measuring the switching current as function of an externally applied flux bias~\cite{Golubov2004,  Della_prl_2007}. The result may, however, demand a comprehensive analysis due to the system dynamics involved.

An alternative approach is to measure the microwave admittance of an RF SQUID and probe the supercurrent flowing through a loop with a single JJ~\cite{Sochnikov2015,NatPhys2017cpr,Haller2022}. This approach offers the added benefit of being able to detect microwave loss, which arises from driving transitions between the ABS~\cite{Dassonneville2013, Ferrier2013, Dassonneville2018, Haller2022}. The method allows to trace the phase-dependent energy gap. ABS with high transmission have smaller transition energies---and thus yield larger contributions to the loss---at phase bias $\pi$. For systems with few ABS, the method allows to resolve transitions between individual pairs of ABS in two-tone spectroscopy measurements~\cite{Janvier2015, vanWoerkom2017, Hays2018, Tosi2019}.

In this article, we report microwave admittance measurements on RF SQUIDs with HgTe QW Josephson junctions. We extract the CPR and phase-sensitively detect the microwave loss. From a comparison with tight-binding simulations we infer the presence of a gap at phase $\pi$ and obtain the general form of the ABS spectrum. Our estimate of the gap in the ABS spectrum matches the values reported previously~\cite{Bocquillon2017}. Our study highlights the versatility of microwave measurements in exploring semiconductor Josephson junctions, emphasizing the critical role of interface properties in the performance of side-contacted Josephson devices.

\section{Devices and experimental techniques}

\begin{figure}
	\centering
	\includegraphics[scale=1]{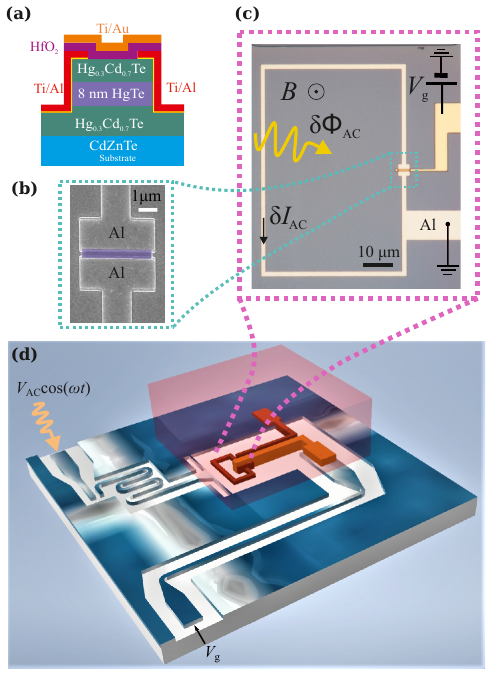}
	\caption{
		\textbf{Microwave admittance measurement of RF SQUID HgTe device.}
		\textbf{(a)} A side-cut of a HgTe Josephson junction layer stack with side-contacts and the $\mathrm{Ti/Au}$ gate deposited on the insulating $\mathrm{HfO_2}$ layer.
		\textbf{(b)} Scanning electron microscopy (SEM) image of Josephson junction in a device, identical to D1, before the gate deposition.
		\textbf{(c)} An optical image of the RF SQUID device~D1 showing the device topology and measurement principle.
		\textbf{(d)} A sketch of the assembled flip-chip device. The RF SQUID on the HgTe chip (pink, top) is attached to the sapphire wafer with a quarter wavelength superconducting resonator (blue, bottom). The resonator is coupled to a transmission line carrying the microwave excitation signal $V_\mathrm{AC}\cos({\omega}t)$. A second line is used to apply the gate voltage $V_\mathrm{g}$.
	}
	\label{fig:device}
\end{figure}
We study {Hg}{Te} quantum well Josephson junctions. Three devices are measured using RF techniques (D1, D2 \& D3), and a reference device is used for DC transport measurements (D4). The junctions are fabricated from an epitaxially-grown, 8-nm-thick HgTe QW layer~[Fig.\,\ref{fig:device}(a)]. The weak links are defined by wet etching a mesa. The sides of the mesa are then argon-milled to remove oxide and contacted with a $\mathrm{Ti/Al/Ti/Au}$ electrode stack [Fig.~\ref{fig:device}(b)] on opposite side-surfaces of the mesa~\cite{Bendias2018}. Here, the bottom 9-$\unit{\nano\meter}$-thick Ti and the top 15-$\unit{\nano\meter}$-thick Au layers are used for adhesion and oxidation protection of the 115-$\unit{\nano\meter}$-thick Al electrode, respectively. (The layer thicknesses given above are for D1 and D2. For the slightly differing parameters of D3 and D4, details of the sample fabrication, and microscope images see Supplemental Material, Note 1.) A $\mathrm{Ti/Au}$ top gate is placed on a thin insulating layer of $\mathrm{HfO_2}$ to control the charge carrier density in the quantum well.

The side-contacted devices, unlike the top-contacted junction geometry, do not involve intermediate transport through a proximitized area of the semiconductor, but rather have the states in semiconductor directly coupled to the superconducting electrodes. This also results in a well-defined length of the ABS trajectories and a preserved directionality of the Andreev reflection processes at the superconductor interfaces.

For RF detection, we embed the HgTe JJ in an RF SQUID loop [Fig.~\ref{fig:device}(c)] that is inductively coupled to a readout resonator, as schematically shown in Fig.~\ref{fig:device}(d). Devices D1 and D2 have a nominal electrode separation $l=500\,\mathrm{nm}$ and a junction width $W=3.8\,\unit{\micro\meter}$.
The RF SQUID loop area is $S_\mathrm{loop}=3290\,\unit{\micro\meter\squared}$. An extra ground connection is added to the loop to define a reference potential for the gate voltage. The third device, D3, is shorter with $l=200\,\unit{\nano\meter}$ and a RF SQUID loop area, $S_\mathrm{loop}=800\,\unit{\micro\meter\squared}$. Device D4 has the same dimensions as D3, but we attached a four-terminal leads pattern for DC measurements. All devices are designed such that the junction length is comparable to the coherence length in the material, i.e., $l \gtrsim \xi = \hbar v_\mathrm F/2 \Delta_0\approx1.94\,\unit{\micro\meter}$~\cite{PhysRevB.46.12573,Ohta1997} where $v_\mathrm F$ is the Fermi velocity and $\Delta_0\approx136\,\unit{\micro\electronvolt}$ is the superconducting gap in the electrodes (details on this estimate are given in \hyperref[sec:dc]{Appendix A}). The data presented in the main text are from measurements on device~D1. Microwave measurements on devices~D2 and D3 are reported in the Supplemental Material, Notes 10 and 11.

For D1, D2 and D3 a $\lambda /4$ microwave readout resonator with unloaded resonance frequency $f_\mathrm{r,0} \approx 4.78\,\unit{\giga\hertz}$ and internal quality factor $Q_\mathrm{i,0}\approx1.8\times10^{5}$ is fabricated from a thin Nb film on a separate sapphire chip. We flip-chip bond the RF SQUID to the readout resonator so that the loop is placed close to the grounded end of the resonator for inductive coupling. The bonding procedure is described in Supplemental Material, Note 2.

To measure the microwave response of the RF SQUID, a microwave excitation, $V_\mathrm{AC}\cos({\omega}t)$, with angular frequency $\omega$ close to $2\pi{f_\mathrm{r,0}}$, is applied to the resonator via the capacitively-coupled transmission line. The oscillating current in the resonator creates a time-dependent AC flux, $\delta \Phi_\mathrm{AC}$, penetrating the RF SQUID loop. The complex microwave admittance (i.e., the inverse of the measured impedance $Z_\mathrm{m}$), $Y_\mathrm{m}=Z_\mathrm{m}^{-1} = \delta I_\mathrm{AC}/(j{\omega}\delta \Phi_\mathrm{AC})$, is read out (Supplemental Material, Note 6) via a shift in the resonance frequency $f_\mathrm{r}$ and a change in the internal quality factor $Q_\mathrm{i}$ of the resonator~\cite{Haller2022}:
\begin{equation}
	\mathrm{Im}[Y_\mathrm{m}] = -\frac{{\pi}Z_\mathrm{r}}{2({\omega}M)^2}(f_\mathrm{r} - f_\mathrm{r,0})/f_\mathrm{r,0}
	\label{eq:admittance_im}
\end{equation}
and
\begin{equation}
	\mathrm{Re}[Y_\mathrm{m}] = \frac{{\pi}Z_\mathrm{r}}{4({\omega}M)^2}(Q_\mathrm{i}^{-1} - Q_\mathrm{i,0}^{-1}).
	\label{eq:admittance_re}
\end{equation}
Here, $Z_\mathrm{r}\approx50\,\unit{\ohm}$ is the characteristic impedance of the transmission line and $M$ is the effective mutual inductance between resonator and RF SQUID loop.

The devices are cooled down in a dry dilution refrigerator with a nominal base temperature below $7\,\mathrm{mK}$. The microwave drive signal is guided to the resonator via a series of attenuators anchored at different temperatures. The reflected signal is passed through a microwave amplification circuit and measured by a network analyzer. A detailed schematic of the measurement setup is given in Supplemental Material, Note 3. We have verified that the microwave excitation on the RF SQUID is small enough to work in the linear regime (Supplemental Material, Note 5).

In addition to the RF excitation, we can apply a static magnetic field $B$ perpendicular to the sample plane using a home-made superconducting coil. The magnetic flux threads through the RF SQUID loop and controls the total phase drop $\varphi_\mathrm{ext} = 2\pi BS_\mathrm{loop}/\Phi_\mathrm{0}$ in the RF SQUID structure, where $\Phi_{0} = h/2e$ is the magnetic flux quantum. To calculate the phase difference between the JJ terminals, $\varphi$, we take into account the circulating current in the RF SQUID loop:
\begin{equation}
	\varphi = \varphi_\mathrm{ext} - \frac{2\pi}{\Phi_0}L_\mathrm{loop}I_\mathrm{J}(\varphi),\label{eq:flux_screening}
\end{equation}
where $L_\mathrm{loop}$ and $I_\mathrm{J}(\varphi)$ are the loop inductance and the Josephson supercurrent, respectively. The resonance frequency  $f_\mathrm{r}(B)$ and the internal quality factor $Q_\mathrm{i}(B)$ are periodic functions in $B$ with the periods $B_{2\pi}{\approx}0.629\,\unit{\micro\tesla}$ for D1-D2 and $B_{2\pi}{\approx}2.58\,\unit{\micro\tesla}$ for D3.

Eq.~\ref{eq:flux_screening} also defines the hysteresis in RF SQUIDs, which stems from the finite inductance $L_\mathrm{loop}$ of the RF SQUID loop. Hysteresis occurs when $\frac{\partial\varphi}{\partial\varphi_\mathrm{ext}}<0$, resulting in jumps of $\varphi$ as a function of applied $\varphi_\mathrm{ext}$. Using Eq.\,\ref{eq:flux_screening}, this condition can be translated to:
\begin{equation}
	L_\mathrm{J}^{-1}(\varphi) < -L_\mathrm{loop}^{-1},
	\label{eq:hysteresis}
\end{equation}
where $L_\mathrm{J}^{-1} = \frac{2\pi}{\Phi_0}\frac{\partial I_\mathrm{J}}{\partial \varphi}$ is the inverse Josephson inductance.

We observe hysteresis in the flux response of the RF SQUID oscillations at large carrier density (i.e., at more positive gate voltage) and monitor its onset to determine the magnitude of $I_\mathrm J$ relative to $L_\mathrm{loop}^{-1}$ using Eq.~\ref{eq:hysteresis} (the analysis is described in Supplemental Material, Note 7). We calculate $L_\mathrm{loop}\approx251\,\unit{\nano\henry}$ for D1, estimating the kinetic inductance contribution using Ref.~\cite{Hu2020} (details and values for other devices are provided in Supplemental Material, Note 8).

When analyzing the measurements, we take into account that the measured impedance is the sum of the complex JJ impedance $Z$ and the impedance of the loop, $Z_\mathrm{m} = Z + j{\omega}L_\mathrm{loop}$. The microwave junction impedance is the reciprocal of the junction admittance, $Y=Z^{-1}$.

\subsection*{Analysis of the microwave admittance}

\begin{figure*}
	\centering
	\includegraphics[scale=1]{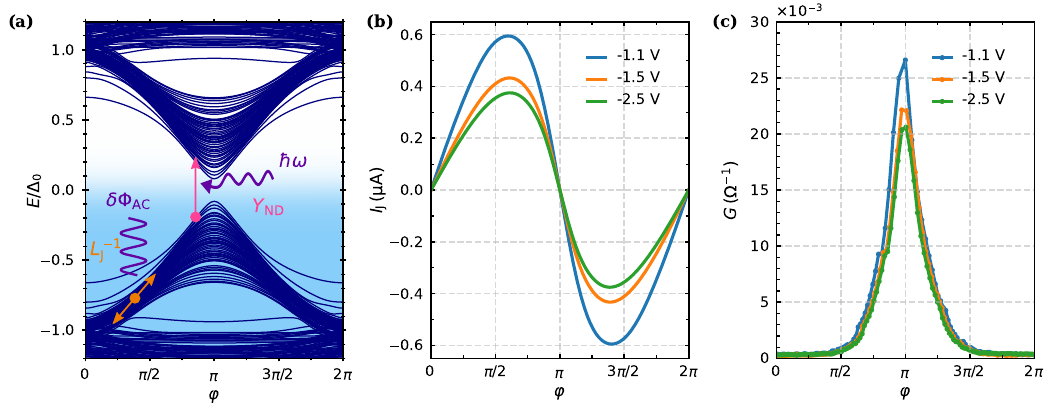}
	\caption{
		\textbf{Phase-dependent microwave response of an RF SQUID.}
		\textbf{(a)} A schematic representation of the physical mechanisms governing the microwave response of the Andreev bound states. The dark blue lines are the ABS spectrum, generated by a tight-binding calculation with JJ geometry of device D1, Fermi velocity $v_\mathrm{F}=8\times10^5\,\unit{\meter/\s}$, Fermi energy $E_\mathrm{F}=10.35\,\unit{\milli\electronvolt}$, and a superconducting gap in the electrodes of $\Delta_0=136\,\unit{\micro\electronvolt}$. In the calculation, we additionally introduced a random disorder potential $E_\mathrm{dis}=3\,\unit{\milli\electronvolt}$ and opened a gap at phase $\pi$ with a $20\,\unit{\nm}$ wide barrier potential of $E_\mathrm{\delta}=8.5\,\unit{\milli\electronvolt}$. At different values of phase bias, we sketch the inductive Josephson response $L_{\mathrm{J}}^{-1}$ arising from the probing flux $\delta\Phi_\mathrm{AC}$ driving the junction (orange, left) and the non-diagonal admittance component $Y_\mathrm{ND}$ related to the microwave-induced transitions in the ABS spectrum caused by photons with energy $\hbar\omega$. The blue background indicates the thermal occupation at low temperature.
		\textbf{(b)} The extracted CPR for device D1 at temperature $T=100\,\unit{\milli\kelvin}$ and gate voltages $V_\mathrm{g}=-1.1,\,-1.5,\,-2.5\,\unit{\volt}$.
		\textbf{(c)} The shunt conductance $G(\varphi)=\mathrm{Re}[Y]$, measured alongside the data in panel (a), represents the microwave loss in the junction. The sharp peaks indicate that the interband energy spacing is close to $\hbar\omega$ only in a small region close to $\varphi=\pi$, while the gap at $\varphi=0$ is much larger than $\hbar\omega$.
	}
	\label{fig:I_chi_phase_Vg}
\end{figure*}

From the Kubo formula, one can show that the microwave admittance of a JJ in linear response can be expressed as a sum of three components~\cite{Ferrier2013}:
\begin{equation}
	Y = (j{\omega}L_\mathrm{J})^{-1} + Y_\mathrm{D} + Y_\mathrm{ND}.
	\label{eq:admittance}
\end{equation}
The Josephson contribution $(j{\omega}L_\mathrm{J})^{-1}$ is purely imaginary and proportional to the inverse Josephson inductance $L_\mathrm{J}^{-1}$~\cite{Dassonneville2013, Ferrier2013, Dassonneville2018, Haller2022}. The diagonal ($Y_\mathrm{D}$) and off-diagonal ($Y_\mathrm{ND}$) admittance components can have both real and imaginary parts.

The diagonal component is given by~\cite{Ferrier2013}:
\begin{equation}
	Y_\mathrm{D}=-\sum_{n}i_n^2\frac{{\partial}f}{\partial\epsilon}\Big|_{\epsilon=\epsilon_n}\frac{\hbar}{\gamma_\mathrm{D} - j\hbar\omega},
	\label{eq:diagonal_admittance}
\end{equation}
where $\gamma_\mathrm{D}$ is the diagonal relaxation rate and $i_n(\varphi)=\frac{2e}{\hbar}\frac{\partial\epsilon_n}{\partial\varphi}$ is the supercurrent contribution of the ABS with index $n$, that is proportional to the derivative of the phase-dependent energy $\epsilon_n(\varphi)$. $f(\epsilon)$ denotes the occupation probability. $Y_\mathrm{D}$ is a non-adiabatic contribution, that appears due to the thermal relaxation of the populations of the Andreev levels and produces a phase dependence that is symmetric with respect to phase $\pi$~\cite{Ferrier2013}. For ordinary ABS the diagonal contribution disappears at $\varphi=0$ and $\varphi=\pi$ due to the vanishing supercurrent contributions.

The non-diagonal contribution is produced by the microwave-induced transitions in the ABS spectrum~\cite{Ferrier2013}:
\begin{equation}
	\begin{split}
		Y_\mathrm{ND} = -\sum_{n,m{\ne}n} \left| J_{nm} \right|^2 & \frac{f(\epsilon_n) - f(\epsilon_m)}{\epsilon_n - \epsilon_m} \times \\ & \times \frac{\hbar}{j(\epsilon_n - \epsilon_m) - j\hbar\omega + \gamma_\mathrm{ND}},
	\end{split}
	\label{eq:kubo_losses}
\end{equation}
where $\gamma_\mathrm{ND}$ is the non-diagonal relaxation rate and $J_{nm}$ are the matrix elements of the current operator.

In our experimental regime the imaginary parts of $Y_\mathrm{D}$ and $Y_\mathrm{ND}$ are much smaller than $(j{\omega}L_\mathrm{J})^{-1}$, allowing to approximate $Y \approx (j{\omega}L_\mathrm{J})^{-1} + G$, where $G$ is the purely real microwave shunt conductance $G(\varphi)=\mathrm{Re}[Y]=\mathrm{Re}[Y_\mathrm{D}+Y_\mathrm{ND}]$. We justify this approximation in \hyperref[sec:appendix_losses_high_t]{Appendix D} after we determine the relaxation rates $\gamma_\mathrm{D}$ and $\gamma_\mathrm{ND}$.

In Fig.~\ref{fig:I_chi_phase_Vg}(a), we summarize the physical processes that contribute to the microwave admittance at low temperatures. In the relevant limit of short, ballistic Josephson junctions, the low-energy Andreev bound states group in two dense bands that lie symmetric with respect to zero energy. The model spectrum is derived from a tight-binding numerical simulation for the geometry of device D1 using physically relevant model parameters. [For details, refer to Section~\ref{sec:discussion} and \hyperref[sec:appendix_tight_binding]{Appendix B}.] The spectrum is gapped, with only few states below the phase-dependent band edge. The ABS energy gap at $\varphi=0$ is comparable to the pairing potential $\Delta_0$ in the contacts.

This behavior is not universal. The level spacing in the ABS spectrum depends on the ratio between the coherence length $\xi$ and the junction dimensions, as well as on the presence of disorder. For wide junctions ($W/\xi\gg1$) with low disorder, the gap vanishes due to the ABS traveling through the junction at various angles (see Supplemental Material of~\cite{PhysRevResearch.3.L032009}). In contrast, for narrow junctions with large contact separation ($L/\xi>1$), the spectrum contains several repetitions of the ABS bands along the energy axis~\cite{PhysRevB.46.12573}. Increasing disorder opens a gap $\epsilon<\Delta_0$ in the ABS spectrum at phase $\varphi=0$ (the so-called "minigap") in both cases~\cite{PhysRevResearch.3.L032009, Ferrier2013}.

We distinguish three contributions to the admittance (cf. Eq.~\ref{eq:admittance}). (i) The Josephson contribution is symbolized by an orange arrow in Fig.~\ref{fig:I_chi_phase_Vg}(a) and accounts for inductive response of the junction to the probing flux $\delta\Phi_\mathrm{AC}$. (ii) The mechanism of non-diagonal admittance is illustrated with a pink arrow. It marks a transition between an occupied state at negative energy and an unoccupied one at positive energy. (iii) The third process is the diagonal admittance, which arises from the relaxation of occupation driven out of equilibrium by the probing AC flux. This process requires $\frac{{\partial}f}{\partial\epsilon}\ne0$ (cf. Eq.~\ref{eq:diagonal_admittance}). In Fig.~\ref{fig:I_chi_phase_Vg}(a) the occupation probability $f$ is illustrated by the background color gradient. At low temperatures the diagonal admittance is strongly suppressed if the gap at phase $\pi$ is large enough. We discuss the influence of $Y_\mathrm{D}$ at higher temperature on our measurement in \hyperref[sec:appendix_losses_high_t]{Appendix D}.

To extract the CPR from the admittance data self-consistently, we use an ansatz composed of sine harmonics up to tenth order\,\cite{Golubov2004}:
\begin{equation}
	I_\mathrm{J}(\varphi) = \sum_{k\ge1} A_k\sin(k\varphi).
	\label{eq:cpr_ansatz}
\end{equation}
The solution is found iteratively\,\cite{Haller2022} by determining sets of parameters $A_k$ to fit $Y_\mathrm m (\varphi_\mathrm{ext})$. After each step, $\varphi$ is recalculated [Eq.~\ref{eq:flux_screening}], and $G(\varphi)$ is reevaluated from $Z_\mathrm{m}$. The process is repeated until convergence is reached (see Supplemental Material, Note 7).

The CPR contains important information about the ABS spectrum. $I_\mathrm{J}(\varphi)$ is given by a sum over the contributions of individual ABS\,\cite{Dassonneville2013, Ferrier2013}:
\begin{equation}
	I_\mathrm{J}(\varphi) =\sum_n f\left[\epsilon_n(\varphi)\right] i_n (\varphi).
	\label{eq:cpr_abs}
\end{equation}
Josephson devices with high-transmission channels feature ABS with a strongly-varying dispersion $\epsilon_\mathrm n (\varphi)$ and thus have a non-sinusoidal CPR for which higher-order harmonics $A_{k>1}$ contribute. The skewness $S$ of the CPR is defined using the phase position of the supercurrent maximum $\varphi_\mathrm{max}$ with respect to $\pi/2$: $S=2(\varphi_\mathrm{max} - \pi/2)/\pi$~\cite{Haller2022,PhysRevB.94.115435}. A strongly forward-skewed CPR with $S>0$ indicates that at least some of the ABS disperse strongly with phase, in contrast to a purely sinusoidal CPR found, e.g., in tunneling junctions~\cite{Golubov2004}.
	
A CPR with $S>0$ can appear both in diffusive~\cite{PhysRevB.66.184513,Haller2022,PhysRevResearch.3.L032009} and ballistic~\cite{Della_prl_2007,Sochnikov2015,NatPhys2017cpr} JJs. Short diffusive junctions with perfectly transmitting interfaces can achieve up to $S=0.255$. A skewness $S=1$ is exhibited by narrow ballistic junctions regardless of the length limit. As a result, while being closely related to the JJ transport regime, the skewness cannot be used as the only criterion to determine in what limit the JJ operates.

For the carrier densities, where the RF SQUID is hysteretic, a small phase range close to $\varphi=\pi$ becomes inaccessible for studies of the JJ response. In this case, we extract the CPR by fitting the JJ admittance $Y(\varphi_\mathrm{ext})$, calculated using $f_\mathrm{r,0}$ extrapolated from measurements without RF SQUID hysteresis. The fit is performed for the five lowest harmonics in Eq.~\ref{eq:cpr_ansatz} (Supplemental Material, Note 7). (Values extracted under this condition are marked by open symbols in our plots.) The extracted harmonics $A_k$, however, become less precise when large parts of the phase response are not available for analysis.

\section{Results}
Figure~\ref{fig:I_chi_phase_Vg}(b) and (c) show the extracted CPR $I_\mathrm{J}(\varphi)$ and microwave shunt conductance $G(\varphi)$ for device D1 at $T = 100\,\unit{\milli\kelvin}$ for three different gate voltages. All three CPR curves in Fig.~\ref{fig:I_chi_phase_Vg}(b) exhibit a pronounced forward skewness $S\approx0.22$ for $V_\mathrm{g}=-2.5\,\unit{\volt},-1.5\,\unit{\volt}$ and $S\approx0.2$ for $V_\mathrm{g}=-1.1\,\unit{\volt}$.

\begin{figure*}
	\centering
	\includegraphics[scale=1]{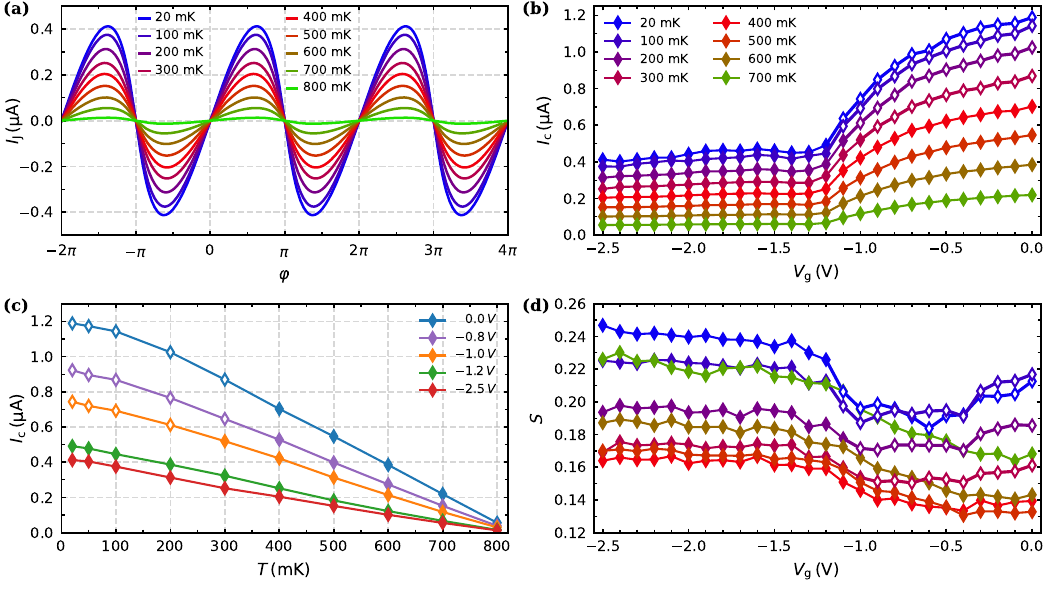}
	\caption{
		\textbf{Current-phase relation (CPR) and its parameters for device D1}
		\textbf{(a)} The CPR traces, measured at gate voltage $V_\mathrm{g}=-2.5\,\unit{\volt}$ and different temperatures (see legend).
		\textbf{(b)} Critical current $I_\mathrm{c}$ at different temperatures{ as a function of gate voltage $V_\mathrm{g}$}, extracted from the data in panel (a) and data at other gate voltages. Empty symbols represent points where RF SQUID hysteresis is present, according to the Eq.~\ref{eq:hysteresis}, and an alternative processing method is used, see main text for further details.
		\textbf{(c)} The same as in (b), shown as a function of temperature $T$ at selected gate voltages (see legend).
		\textbf{(d)} Skewness of the current-phase relation $S$ measured in the same run as $I_\mathrm{c}$ in panel (b). Empty symbols on the right are less precise, since part of the current-phase relation can not be measured there.
	}
	\label{fig:cpr}
\end{figure*}

From the CPR at maximal depletion, $V_\mathrm{g}=-2.5\,\unit{\volt}$, we extract the critical current $I_\mathrm{c}\approx0.38\,\unit{\micro\ampere}$ by reading off the maximal value of $I_\mathrm{J}(\varphi)$. The critical current remains large over the entire gate voltage range, and at $V_\mathrm{g}=0$ we extract $I_\mathrm{c}\approx1.2\,\unit{\micro\ampere}$. These values are significantly larger than the edge supercurrent reported previously \cite{Hart2014, Bocquillon2017}. By comparing $I_\mathrm c$ to an estimate of the supercurrent a single ABS carries, $i_\mathrm{est} \sim \frac{e\Delta_0}{\hbar}\approx33\,\mathrm{nA}$, we conclude that a substantial number of bulk ABS contributes to supercurrent transport, hence a finite bulk carrier density persists down to the lowest values of $V_\mathrm g$, and is even higher at $V_\mathrm{g}=0$.

The presence of large carrier densities in our samples is corroborated by magnetoresistance measurements on device D4 (\hyperref[sec:dc]{Appendix A}), for which the carrier density is found to be $n_\mathrm{e}\sim 10^{12}/\unit{\centi\meter}^2$ in the relevant gating range, see Supplemental Material, Note 4. For the other devices we estimate $n_\mathrm{e}$ based on the gating efficiency~\cite{Fuchs_2024,Liang_2024} and $I_\mathrm{c}$ to be in the mid- to high $10^{11}/\unit{\centi\meter}^{2}$. These values are remarkably higher than the ones extracted from characterizing the starting material (Supplemental Material, Note 1). The effect is strong for devices much shorter than $1\,\unit{\micro\meter}$ and was traced to originate from a short ion-milling treatment used for oxide removal prior to contact deposition.

Any observation of edge state contributions in the junction transport requires the bulk carrier density to be tuned close to zero, either by taking undoped material or by applying a large enough negative $V_\mathrm{g}$. However, in HgTe quantum wells the gate voltage effect is limited by charging/discharging of the states on semiconductor-insulator interface~\cite{Hinz2006} to ${\delta}n_\mathrm{e}\approx1.2\times10^{12}\,\mathrm{cm}^{-2}$. In the case when the initial carrier density exceeds this value, the topological regime remains inaccessible regardless of the applied gate voltage (\hyperref[sec:dc]{Appendix A}).

We can learn more about the ABS spectrum by examining the microwave loss $G(\varphi)$ in Fig.~\ref{fig:I_chi_phase_Vg}(c). At low temperatures, it exhibits sharp peaks at phases corresponding to odd multiples of $\pi$, $\varphi_\mathrm{odd}\equiv(2n+1)\pi$, and it is almost zero at even multiples of $\pi$, $\varphi_\mathrm{even}\equiv2n\pi$. The microwave dissipation originates from photon-induced transitions between Andreev levels~\cite{Dassonneville2013,Ferrier2013,Dassonneville2018, Haller2022}. At the lowest temperatures, the lower (upper) ABS levels are fully occupied (unoccupied). Contributions to $G(\varphi)$ mainly arise from interband transitions across the spectral gap [process (ii) in Fig.~\ref{fig:I_chi_phase_Vg}(a)] when the energy difference $|\epsilon_n - \epsilon_m|$ between the participating bands $n$ and $m$ is close to the microwave photon energy $\hbar\omega=h{f_\mathrm{r}}=19.8\,\unit{\micro\electronvolt}$. Thus peaks in $G(\varphi)$ appear where the gap in the ABS spectrum $E_\mathrm g(\varphi)$ becomes comparable to $\hbar \omega$. Conversely, the vanishing of the dissipation around $\varphi_\mathrm{even}$ indicates that $E_\mathrm g \gg \hbar \omega$ there~\cite{Ferrier2013}. This large change in $E_\mathrm{g}(\varphi)$ combined with the prominent forward skewness of the CPR points to the presence of strongly dispersing ABS, whose contributions dominate in the supercurrent transport~\cite{Sochnikov2015, Murani2017}.

\subsection*{Temperature dependence of microwave response}

We also perform microwave admittance measurements at higher temperatures up to $T=800\,\unit{\milli\kelvin}$. (Supercurrent and phase response of the RF SQUID disappears at $T \approx 850\,\unit{\milli\kelvin}$.) In Fig.~\ref{fig:cpr}(a), we show CPR traces for $V_\mathrm{g}=-2.5\,\unit{\volt}$. The amplitude of $I_\mathrm{J}(\varphi)$ decreases with increasing temperature while the CPR remains forward-skewed. To further investigate this, we extract $I_\mathrm{c}$ [Fig.~\ref{fig:cpr}(b,c)] and $S$ [Fig.~\ref{fig:cpr}(d)] at all gate voltages and temperatures. The points affected by RF SQUID hysteresis, according to Eq.~\ref{eq:hysteresis}, are indicated by the empty symbols.

\begin{figure}
	\centering
	\includegraphics[scale=1]{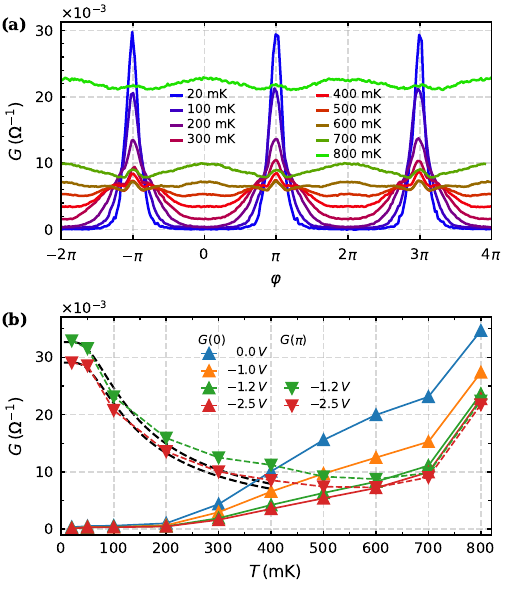}
	\caption{
		\textbf{Temperature-dependent microwave response of the RF SQUID.}
		\textbf{(a)} The shunt conductance $G(\varphi)$, representing microwave loss in the junction, measured alongside the data in \ref{fig:cpr}(a). The appearance of loss at $\varphi=0$ for higher $T$ is related to the thermal broadening reaching the size of the gap between the ABS, allowing intraband transitions in the upper and lower ABS branches.
		\textbf{(b)} The shunt conductance at phase $\varphi=0$ (upward triangles) and $\pi$ (downward triangles). The symbol colors correspond to the same $V_\mathrm{g}$ values as in panel (a). Phase $\varphi=\pi$ is not accessible for $V_\mathrm{g} \geq -1\,\unit{\volt}$ due to RF SQUID hysteresis.
	}
	\label{fig:losses}
\end{figure}

\begin{figure}
	\centering
	\includegraphics[scale=1]{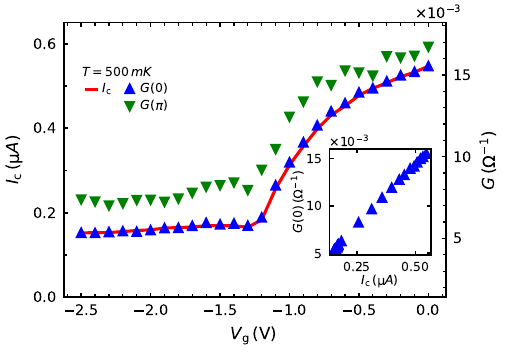}
	\caption{
		\textbf{Comparison of the gate voltage dependencies of critical current and microwave shunt conductance at phase bias $\varphi=0$ and $\pi$.}
		The gate voltage $V_\mathrm{g}$ dependence of critical current $I_\mathrm{c}$ (red lines, correspond to the left axes) and shunt conductance $G$ (symbols, correspond to the right axis), extracted from the measurement at temperature $T=500\,\unit{\milli\kelvin}$ for device D1. By normalizing and offsetting $G(0)$, we perfectly match the $V_\mathrm{g}$ dependence with that of $I_\mathrm{c}$. The inset shows $G(0)$ as a function of $I_\mathrm{c}$ to demonstrate the linear relationship.
	}
	\label{fig:a_k_g_vs_v_g}
\end{figure}

The temperature dependence of $I_\mathrm{c}$ is shown in Fig.~\ref{fig:cpr}(c) for several values of $V_\mathrm{g}$. It is approximately linear at higher temperatures, consistent with calculations for short junctions with high interface transparencies (cf. Ref.\,\cite{Chrestin1994}). It is also compatible with the prediction of Kulik and Omelyanchuk for short ballistic junctions~\cite{Golubov2004}.

The skewness in Fig.~\ref{fig:cpr}(d) only varies by about $25\%$ with $V_\mathrm{g}$ at any temperature, and $S>0.1$ throughout. At $T=20\,\unit{\milli\kelvin}$, the CPR skewness is between $0.2$ and $0.25$. Values in this range are expected for both diffusive junctions with clean interfaces or wide ballistic junctions with non-ideal interfaces. The skewness has a non-monotonic temperature dependence with a minimum close to $T\approx400\,\unit{\milli\kelvin}$. The initial drop relates to the broadening of the thermal distribution, leading to the depopulation of ABS. The increase at higher temperature may be attributed to the contribution of diagonal transitions to the admittance $\mathrm{Im}[Y_\mathrm{D}]$, appearing at higher temperatures and affecting the extraction of the CPR (\hyperref[sec:appendix_losses_high_t]{Appendix D}). This limits the validity of the extracted $S$ values for $T>400\,\unit{\milli\kelvin}$, but has only small effect on the supercurrent magnitude in Fig.~\ref{fig:cpr}(b-c).

The shape of the microwave loss evolves as the temperature is increased, see Fig.~\ref{fig:losses}(a) for data at $V_\mathrm{g}=-2.5\,\unit{\volt}$. First, the peaks at $\varphi_\mathrm{odd}$ broaden. At $T>200\,\unit{\milli\kelvin}$, finite microwave loss is observed for all values of $\varphi$. Eventually, the loss at $\varphi_\mathrm{even}$ exceed that at $\varphi_\mathrm{odd}$ when $T\approx 700\,\unit{\milli\kelvin}$. This can be understood as resulting from ABS with $\epsilon\gg\hbar\omega$ ($\epsilon\ll\hbar\omega$) becoming partially occupied (vacant) at elevated temperatures, causing the rate of interband transitions to decrease. Conversely, intraband transitions (e.g., both $\epsilon_n,\, \epsilon_m > 0$) start to appear and contribute toward the dissipation. Additionally, when $T$ is close to the critical temperature in the leads, $T_\mathrm{c}\approx945\,\unit{\milli\kelvin}$ (extracted from the DC measurements in \hyperref[sec:dc]{Appendix A}), the pairing potential diminishes thus leading to an overall increase in dissipation as seen for the trace at $T = 800\,\unit{\milli\kelvin}$.

In Fig.~\ref{fig:losses}(b), we show the temperature dependence of $G(\varphi)$ at phase bias $\varphi=0$ and $\pi$. Phase $\varphi = \pi$ is not accessible for $V_\mathrm{g}\ge-1\,\unit{\volt}$ and low temperatures due to RF SQUID hysteresis. We observe that the temperature dependence of $G(\pi)$ is non-monotonous with a steep upturn above $700\,\unit{\milli\kelvin}$. This feature corresponds to an nearly uniform increase in the loss for all values of $\varphi$ before the RF SQUID response vanishes at $T \approx 850\,\unit{\milli\kelvin}$.

\begin{figure*}
	\centering
	\includegraphics[scale=1]{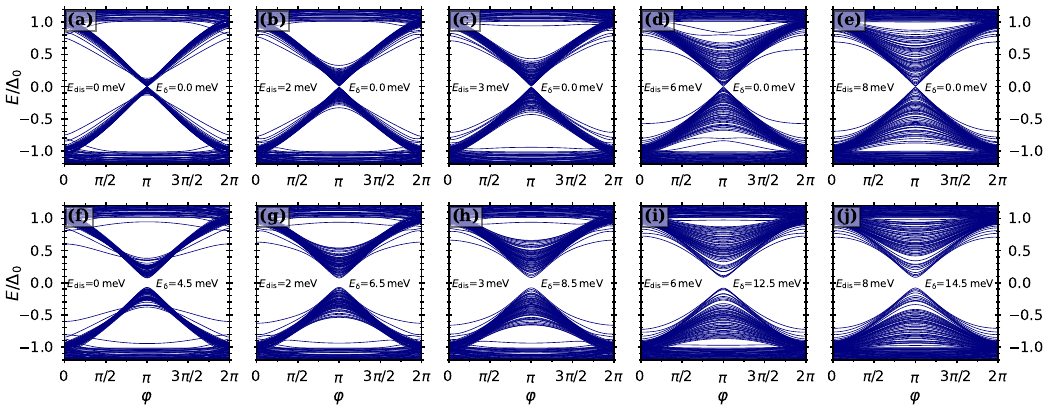}
	\caption{
		\textbf{ABS spectra generated by the tight-binding simulation for a Josephson junction with length $l=500\,\unit{\nano\meter}$ and width $W=4\,\unit{\micro\meter}$, similar to our device D1.} For the simulations, we choose the following parameters: Fermi velocity $v_\mathrm{F}=8\times10^{5}\,\unit{\meter/\second}$, Fermi energy $E_\mathrm{F}=10.35\,\unit{\milli\electronvolt}$, and superconducting gap $\Delta_0=0.136\,\unit{\milli\electronvolt}$.
		\textbf{(a-e)} ABS energies in units of $\Delta_0$ as a function of superconducting phase $\varphi$.
		\textbf{(f-j)} Same as (a-e), but with a potential barrier of strength $E_\mathrm{\delta}$ included and tuned to open a gap $E_\mathrm{g}(\pi)\approx10\,\unit{\micro\electronvolt}$.
	}
	\label{fig:tight_binging_spectra}
\end{figure*}

A common feature in the $G(0)$ traces is the onset of noticeable loss above $200\,\unit{\milli\kelvin}$. Since, at low temperature microwave loss arises from interband transitions, we may interpret this observation as the thermal broadening becoming comparable to the size of the gap between the ABS bands, $E_\mathrm{g}(\varphi)$\,\cite{Ferrier2013,Dassonneville2018}. Hence, $E_\mathrm g (0) \gtrsim 4k_\mathrm B \times 200\,\unit{\milli\kelvin} \approx 70\,\unit{\micro\electronvolt}$, where $k_\mathrm B$ denotes the Boltzmann constant.

At still higher temperatures the loss at $\varphi_\mathrm{even}$ becomes prominent. In Fig.~\ref{fig:a_k_g_vs_v_g}(a), the gate dependencies of $G(0)$ and $G(\pi)$ are shown for $T=500\,\unit{\milli\kelvin}$. We observe that the loss increases with critical current; the dependence can be described by a linear scaling relation, $b G=(I_\mathrm{c}+a)$, with offset parameter $a\approx50\,\unit{\nano\ampere}$ and scaling coefficient $b=38.7\,\unit{\micro\volt}$. This relation holds for a wide range of temperatures [cf. Supplemental Material, Note 9]. Here, $a$ only depends weakly on temperature, whereas $b$ relates to the photon energy and the thermal broadening. The observed scaling of $I_\mathrm{c}$ and $G(0)$ may indicate that the junction transport regime remains the same for all studied gate voltages.

\section{Discussion}
\label{sec:discussion}

One expects that the properties of our side-contacted Josephson junctions are strongly influenced by the interfaces between the superconductor and semiconductor. Moreover, the dissimilarity of the Fermi surfaces of the HgTe QW and the aluminum contacts, along with the possibility of additional disorder and potential barriers at the interfaces, introduces multiple extra parameters into the quantum-mechanical description of the system. It is impractical to cover this vast parameter space by brute force numerical simulation of the microwave response. We therefore settle on developing an intuition on how disorder and potential barriers influence the ABS spectrum in the relevant limits for our devices.  For this purpose, we have done numerical calculations based on a tight-binding model of the Josephson junction with parameters corresponding to device D1.

Our tight-binding model is based on the Bogoliubov-de Gennes equations for a parabolic electron band, as is appropriate for the high carrier density in our devices. The Fermi level is set to $E_\mathrm{F}=10.35\,\unit{\milli\electronvolt}$, ensuring $E_\mathrm{F}\gg\Delta_0$, while the effective mass is adjusted to reproduce the Fermi velocity $v_\mathrm{F}=8\times10^{5}\,\unit{\meter}/\unit{\second}$. This regime corresponds to a Fermi wavevector of $k_\mathrm{F}\approx0.0393\,\unit{\nano\meter^{-1}}$, which is smaller than expected for the carrier density of our devices, yet results in matching magnitudes for $I_\mathrm{c}$ and normal state resistance. The aspect ratio of the junction $L/W=1/8$, is chosen to match the dimensions of device D1. To introduce disorder, a random electrostatic potential uniformly distributed between $-E_\mathrm{dis}$ and $E_\mathrm{dis}$ is applied to the normal region, where $E_\mathrm{dis}$ represents the disorder magnitude, which we vary. A potential barrier with magnitude $E_\delta$ is added to a section in the middle of the junction. Setting $E_\delta>0$ allows us to simulate non-ideal interfaces. Further details of the tight-binding model and the reasons for our choice of $E_\mathrm{F}$ are provided in \hyperref[sec:appendix_tight_binding]{Appendix B}.

First we study the effect of disorder on a junction with transparent interfaces ($E_\delta=0$). In Fig.~\ref{fig:tight_binging_spectra}(a-e) we show the ABS spectra [$E(\varphi)$] generated for the geometry of device D1 with different disorder magnitude, varied from absent, $E_\mathrm{dis}=0$, to strong, where $E_\mathrm{dis}=8\,\unit{\milli\electronvolt}$, close to the Fermi energy $E_\mathrm{F}=10.35\,\unit{\milli\electronvolt}$. With $E_\mathrm{dis}=0$ the spectrum is similar to that of a ballistic 1D contact with a slight variation in energy, producing a gap at phase 0 close to $\Delta_0$. With increasing disorder the ABS band broadens strongly at phase $\varphi=\pi$, while at phase $\varphi=0$ the states still bunch at $E_\mathrm{g}(0)\sim{\Delta_0}$. This behavior is a consequence of the coherence length $\xi = \hbar v_\mathrm F/2 \Delta_0 \approx1.94\,\unit{\micro\meter}$ not only being well above the junction length $L$, which allows only one band of ABS to be present in the spectrum, but also comparable to the junction width $W$, which reduces the spread at $\varphi=0$ of ABS in the band.

Next, we discuss the effect of finite interface transparency. For highly transmissive barriers, the essential features are captured by introducing a single potential barrier inside the JJ~\cite{PhysRevB.46.12573} [Fig.~\ref{fig:tight_binging_spectra}(f-j).] This opens a gap $E_\mathrm{g}(\pi)$ at phase $\pi$. For lower transparency, however, backscattering at the S-TI interfaces leads to transmission resonances. We do not observe such resonances in the experiment and limit our discussion to the former case. In Fig.~\ref{fig:tight_binging_spectra}(f-j) we present the spectra generated with the same values of $E_\mathrm{dis}$ as in Fig.~\ref{fig:tight_binging_spectra}(a-e), but with barrier strength $E_\mathrm{\delta}$ adjusted to obtain $E_\mathrm{g}(\pi)\approx10\,\unit{\micro\electronvolt}$. We demonstrate below that this value results in the best match with our experimental data.

Using the tight-binding spectra shown in Fig.~\ref{fig:tight_binging_spectra}, the microwave admittance $Y$ can be calculated following Eqs.~\ref{eq:admittance}-\ref{eq:kubo_losses}. However, before comparing the full phase-dependent microwave response from the simulation with experimental data, we specifically focus on the microwave loss at low temperatures and a phase difference of $\pi$. At this phase the diagonal contribution $Y_\mathrm{D}$ vanishes, and a simplified analytical expression can be obtained for the non-diagonal contribution $Y_\mathrm{ND}$.

In Eq.~\ref{eq:kubo_losses} both the energies of ABSs $\epsilon_n,\epsilon_m$ and the matrix elements of the current operator $J_{nm}$ are phase-dependent, which complicates direct analysis. At phase $\varphi=\pi$, the situation is different because the matrix elements for symmetric particle-hole states $J_{n,-n}$ become dominant and, according to our simulations (\hyperref[sec:appendix_matrix_elements]{Appendix C}), the magnitude of these elements for the lowest lying ABS is close to the maximum supercurrent of the state $|J_{n,-n}(\pi)|{\approx}i_\mathrm{max}\sim\frac{e\Delta_0}{\hbar}$. Applying this simplification, the loss at phase $\pi$ can be rewritten as:
\begin{equation}
	\begin{split}
		& G_\mathrm{ND}(\pi)\approx i^2_\mathrm{max}\hbar\gamma_\mathrm{ND}\sum_{n>0} \frac{1 - 2 f(\epsilon_n,T)}{2\epsilon_n}\times
		\\ & \times \bigg[\frac{1}{(2\epsilon_n + \hbar\omega)^2 + \gamma_\mathrm{ND}^2} + \frac{1}{(2\epsilon_n - \hbar\omega)^2 + \gamma_\mathrm{ND}^2}\bigg].
	\end{split}
	\label{eq:model_losses}
\end{equation}

According to Eq.~\ref{eq:model_losses}, the temperature dependence of $G_\mathrm{ND}(\pi)$ arises from the factors $1 - 2 f(\varepsilon_n,T)$ inside the sum. The result of the sum strongly depends on the presence and size of a gap in the ABS spectrum at $\varphi=\pi$. In the absence of a gap, the states with smallest $\varepsilon_n$ dominate in the sum, due to the factor $1/\varepsilon_n$. This leads to a strong temperature dependence with a sharp decrease at low temperatures. If a gap is present, we can approximate $G(\pi)\approx G_\mathrm{ND}(\pi) \propto [1 - 2 f(E_\mathrm{eff},T)]$ where $E_\mathrm{eff} > E_\mathrm{g}(\pi)$ is the mean transition energy for transitions between the lower and upper ABS band. This produces a plateau at low temperatures, comparable to the data in Fig.~\ref{fig:losses}(b). We prove this consideration by generating temperature dependence of full $\mathrm{Re}[Y]$ from our numerical simulations in \hyperref[sec:appendix_matrix_elements]{Appendix C}, where we also observe that the actual difference between $E_\mathrm{eff}$ and $E_\mathrm{g}$ is typically small, but depends on the amount of disorder and $\gamma_\mathrm{ND}$.

\begin{figure*}
	\centering
	\includegraphics[scale=1]{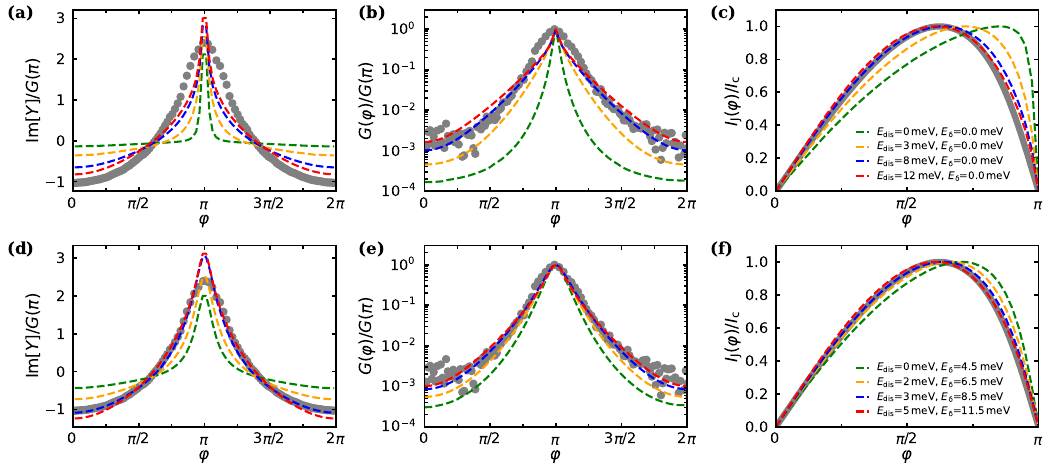}
	\caption{
		\textbf{Comparison of microwave response generated by tight-binding calculation and experimental data.} The gray symbols and gray solid lines are the experimental data for device D1 at gate voltage $V_\mathrm{g}=-2.5\,\unit{\volt}$ and temperature $T=20\,\unit{\milli\kelvin}$. \textbf{(a)} Imaginary part of microwave admittance, $\mathrm{Im}[Y]$, calculated for several magnitudes of the disorder potential $E_\mathrm{dis}$ without a potential barrier $E_\mathrm{\delta}=0$ at $T=20\,\unit{\milli\kelvin}$ [legend in panel (c)]. The values are normalized by the microwave shunt conductance at phase $\pi$, $G(\pi)$. The relaxation rates are taken as $\gamma_\mathrm{ND}=\gamma_\mathrm{D}=100\,\unit{\micro\electronvolt}$. \textbf{(b)} Real part of the microwave admittance (microwave shunt conductance), $G$, for the same parameters as (a), normalized by the value at phase $\pi$. \textbf{(c)} Supercurrent, normalized by the critical current, $I_\mathrm{c}$, for the same parameters as (a). \textbf{(d-f)} Same as (a-c) but with $E_\mathrm{\delta}$ opening a gap at phase $\pi$, $E_\mathrm{g}\approx10\,\unit{\micro\electronvolt}$ [legend in panel (f)]. The relaxation rates are taken as $\gamma_\mathrm{ND}=\gamma_\mathrm{D}=40\,\unit{\micro\electronvolt}$.
	}
	\label{fig:admittance_compare}
\end{figure*}

This can approximate the low-temperature loss at $\pi$ as $G(\pi) \approx G_\mathrm{ND}(\pi) \propto [1 - 2 f(E_\mathrm{eff},T)]$. Such fits are indicated as the dashed black lines in Fig.~\ref{fig:losses}(b) and yield $E_\mathrm{eff}=17\,\unit{\micro\electronvolt}$ with reasonable accuracy, which matches results of our simulation with a notable gap $E_\mathrm{g}(\pi)\sim10\,\unit{\micro\electronvolt}$.

To further evidence the presence of a gap $E_\mathrm{g}(\pi)$ in our devices, we simulate the full microwave response using a tight-binding model for both $E_\mathrm{\delta}=0$ [Fig.~\ref{fig:admittance_compare}(a-c)] and $E_\mathrm{\delta}\ne0$  [Fig.~\ref{fig:admittance_compare}(d-f)], producing a gap $E_\mathrm{g}\approx10\,\unit{\micro\electronvolt}$ at $T=20\,\unit{\milli\kelvin}$. We compare these simulations (dashed lines) with the shape and relative magnitude of the admittance components from our measurements (gray dots and lines). We take the data measured for device~D1 at $V_\mathrm{g}=-2.5\,\unit{\volt}$, where the whole phase range is accessible. By varying the simulation parameters we first try to match the CPR and $\mathrm{Im}[Y]$, as the CPR does not depend and $\mathrm{Im}[Y]$ only weakly depends on $\gamma_\mathrm{ND}$ in the relevant parameter range. After that we adjust $\gamma_\mathrm{ND}$ to match $G=\mathrm{Re}[Y]$. As the diagonal contribution is small at $T=20\,\unit{\milli\kelvin}$ we take $\gamma_\mathrm{D}=\gamma_\mathrm{ND}$.

We start with the case $E_\mathrm{\delta}=0$. For a good match with the CPR [Fig.~\ref{fig:admittance_compare}(c)], we have to introduce a large disorder, $E_\mathrm{dis}\in[8,12]\,\unit{\milli\electronvolt}$. In this case, the microwave shunt conductance [Fig.~\ref{fig:admittance_compare}(b)] exhibits a much sharper phase dependence than the experimental data, even after optimizing $\gamma_\mathrm{ND}=\gamma_\mathrm{D}=100\,\unit{\micro\electronvolt}$ to achieve the best match for $G(\varphi)/G(\pi)$. Furthermore, the imaginary part of the microwave admittance in Fig.~\ref{fig:admittance_compare}(a) shows significant discrepancies from the experimental data, both in its shape and in its ratio to $G(\pi)$. A further increase of $E_\mathrm{dis}$ also does not lead to a better fit of the measured admittance.

For comparison, we present a modeling with $E_\mathrm{g}(\pi)\approx10\,\unit{\micro\electronvolt}$ and $E_\mathrm{dis}\in[3,5]\,\unit{\milli\electronvolt}$, which aligns much better with the experimental CPR data [Fig.~\ref{fig:admittance_compare}(f)]. We achieve a much closer match to the experimental ratio $G(\varphi)/G(\pi)$ in Fig.~\ref{fig:admittance_compare}(e) with $\gamma_\mathrm{ND}=\gamma_\mathrm{D}\approx40\,\unit{\micro\electronvolt}$ (results for different values are shown in \hyperref[sec:appendix_losses_high_t]{Appendix D}). Similarly, for $\mathrm{Im}[Y]$ in Fig.~\ref{fig:admittance_compare}(d), we obtain the best agreement using $E_\mathrm{dis}=3\,\unit{\milli\electronvolt}$ and $E_\mathrm{\delta}=8.5\,\unit{\milli\electronvolt}$. Further variation of these parameters does not yield a better match with both admittance components simultaneously.

The comparisons between experiment and modeling in Fig.~\ref{fig:admittance_compare}, together with the temperature dependencies shown in Fig.~\ref{fig:losses}(b), provide compelling evidence for the opening of a gap at phase $\pi$, and highlights the significant role of the interfaces in the transport properties of our Josephson junction devices. The mismatch in the peak of $\mathrm{Im}[Y]$ may be attributed to an imprecise estimate of the superconducting gap $\Delta_0$ or, more likely, to the simplicity of the interface model used, which treats it as a barrier. Thus further improvements in modeling may require a more sophisticated treatment of the interface effects.

In \hyperref[sec:appendix_losses_high_t]{Appendix D}, we use the model with $E_\mathrm{dis}=3\,\unit{\milli\electronvolt}$ and $E_\mathrm{\delta}=8.5\,\unit{\milli\electronvolt}$ to calculate the microwave response at higher temperatures. In this analysis, we observe reasonable agreement below $400\,\unit{\milli\kelvin}$, with the exception from the symmetric shoulders in $G$ that stem from the diagonal component of the admittance. Additionally, in \hyperref[sec:appendix_losses_high_t]{Appendix D} we justify our choice of $\gamma_\mathrm{ND}$ and demonstrate that the imaginary components of $Y_\mathrm{D}$ and $Y_\mathrm{ND}$ are sufficiently small to ensure the validity of our CPR extraction procedure, confirming that the adiabatic approximation holds.

We finally note that several theoretical works suggest that the existence of a (sharp) dissipation peak at phase $\pi$ can be interpreted as a signature of topological Andreev states~\cite{Dmytruk2016, Murani2017PhysRevB, Trif2018}. However, we demonstrate that  strongly dispersing bulk states produce a similar peak which thus cannot be interpreted as a definitive topological signature. Temperature-dependent peak broadening, as observed in Ref.~\cite{murani2019microwave}, could provide additional support for topological interpretation, but further evidence is needed to definitively confirm it.

\section{Summary}

We have measured the microwave response of RF SQUIDs based on HgTe QW Josephson junctions. From this response, we have extracted the CPR and the phase-dependent microwave loss of the junction and compared the measured microwave admittance to a tight-binding simulation for the relevant parameters. Our analysis, including the temperature dependence of the microwave loss, confirms the existence of a small gap at phase $\pi$ in the ABS spectrum. The magnitude of this gap corresponds to high, yet not unitary, transparency of the interfaces. Crucially, without the inclusion of this gap, the experimental data cannot be accurately reproduced by the model.

The carrier density in our devices allowed us to explore only the regime with dominating bulk supercurrent, where the topological effects are hidden and cannot be isolated. The increase likely originates from unintentional doping by ion-milling the interfaces of HgTe mesa before electrode deposition. This highlights the importance of gentle interface treatment in the fabrication of JJ devices. Future work will focus on samples with lower bulk carrier density that can be reliably tuned into the topological regime, allowing us to perform the spectroscopy of the Andreev bound states in this limit.

We conclude that the microwave admittance measurements provide a useful way to extract the JJs properties and could be widely adopted in other JJ platforms~\cite{Pribiag2015, Banerjee2022, Moehle2022}.

\begin{figure*}
	\centering
	\includegraphics[scale=1.0]{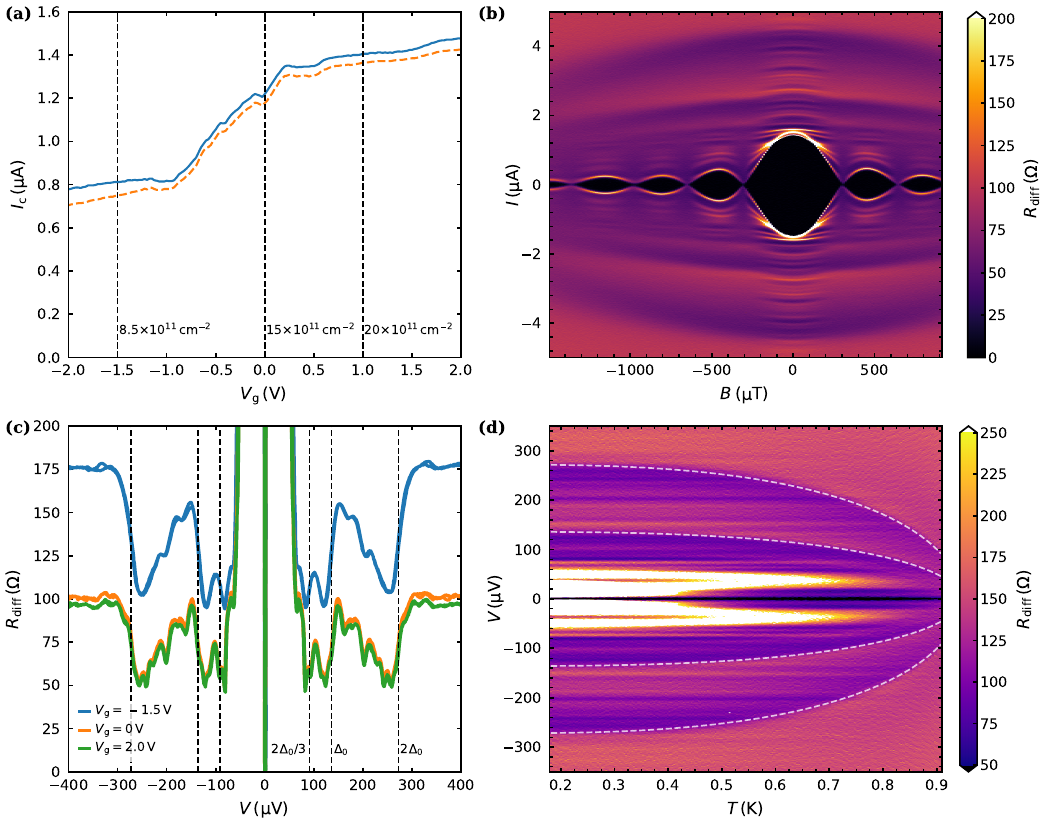}
	\caption{\textbf{Measurements on the complementary (DC) device D4.} \textbf{(a)} Critical current (blue, solid) and retrapping current (orange, dashed) at $T=34\,\unit{\milli\kelvin}$ as a function of gate voltage, extracted from DC measurements. Vertical lines indicate the gate voltages where the carrier density was extracted from magnetoresistance measurements. (b) DC measurements in magnetic field at $V_\mathrm{g}=0\,\unit{\volt}$. The differential resistance is extracted from numerically differentiating the I-V curves. (c) Differential resistance as a function of voltage bias at three different gate voltages. Dashed lines indicate the positions of the MAR features. (d) The temperature dependence of the differential resistance at $V_\mathrm{g}=-1.5\,\unit{\volt}$. The white dashed lines indicate the result of the interpolation formula for $\pm\Delta$ and $\pm2\Delta$.}
	\label{fig:dc_measurements}
\end{figure*}

\begin{acknowledgments}
We thank G. A. Steele for the insightful discussions and R. S. Deacon for his advice on building the RF measurement setup and designing the microwave resonators. We gratefully acknowledge the financial support of the ERC-Advanced Grant Program (project ``4-TOPS'', grant agreement No 741734), the Free State of Bavaria (the Institute for Topological Insulators), and the Deutsche Forschungsgemeinschaft (DFG, German Research Foundation) – SFB 1170 (Project-ID 258499086) and EXC2147 "ct.qmat" (Project‐ID 39085490).

W.L. and S.U.P. contributed equally to this work.
\end{acknowledgments}

\section*{Appendix A: DC measurements on device D4}
\label{sec:dc}

In this appendix, we present basic DC transport measurements on a side-contacted Josephson junction (D4), fabricated using the same technology as RF SQUID device (D1-D3). DC measurements were performed in a different experimental setup, using an Oxford Kelvinox 400 dilution refrigerator, with base temperature $T=34\,\unit{\milli\kelvin}$.

In Fig.~\ref{fig:dc_measurements}(a) we show the critical current, $I_\mathrm{c}$, (blue line) and retrapping current (orange line) extracted from I-V curves measured at varied gate voltage $V_\mathrm{g}$. The product of critical current and normal state resistance is $I_\mathrm{c}R_\mathrm{N}\approx120\,\unit{\micro\volt}$ throughout. At the positions marked by dashed lines we extract the carrier density $n_\mathrm{e}$ from magnetoresistance measurements in a separate experiment. At gate voltage $V_\mathrm{g}=0$, the extracted $n_\mathrm{e}=1.5\times10^{12}\,\mathrm{cm}^{-2}$ is almost an order of magnitude larger than the value obtained from analyzing a macroscopic Hall bar from the same wafer. By applying a gate voltage to D4, we can tune $n_\mathrm{e}$ in the range $n_e=0.8\times10^{12}-2\times10^{12}\,\unit{\centi\meter}^{-2}$. This matches the typical hysteresis-free gate action in HgTe QWs of ${\delta}n_\mathrm{e}\approx1.2\times10^{12}\,\mathrm{cm}^{-2}$~\cite{Hinz2006}. At the same time, the critical current $I_\mathrm{c}$ changes between $I_\mathrm{c}=1.5\,\unit{\micro\ampere}$ at $V_\mathrm{g}=2\,\unit{\volt}$ and $I_\mathrm{c}=0.75\,\unit{\micro\ampere}$ at $V_\mathrm{g}=-2\,\unit{\volt}$, saturating alongside the $n_\mathrm{e}$. Later investigations (not reported here) support the assumption, that in small HgTe devices the carrier density is increased by Ar ion-milling the contact areas prior to metal deposition. We expect a similar behavior for D1-D3, where $n_\mathrm{e}$ could not be assessed directly. In the next generation of devices, where the step has been excluded by shifting to a different process technology, low carrier density has been observed~\cite{liu2024perioddoubling}.

In the above mentioned range of carrier densities, we investigate the regime for which a large number of bulk ABS participate in supercurrent transport. We can also study the I-V curves above $I_\mathrm{c}$. First we investigate the magnetic field dependence at $V_\mathrm{g}=0$ in Fig.~\ref{fig:dc_measurements}(b) where we show extracted the differential resistance $R_\mathrm{diff}=dV/dI$. We observe a typical interference pattern for $I_\mathrm{c}$ with three large $R_\mathrm{diff}$ features, that scale together in magnetic field, as well as a number of other small-period oscillations. From analyzing the differential resistance from the I-V curves at different $V_\mathrm{g}$ in Fig.~\ref{fig:dc_measurements}(c), we find that the large features do not move with gate voltage (i.e., carrier density) and can be assigned to a multiple Andreev reflection sequence $2\Delta_0$, $\Delta_0$ and $2\Delta_0/3$ with $\Delta_0\approx136\,\unit{\micro\electronvolt}$. This value matches the expected gap suppression for a contact stack, consisting partially of non-superconducting Au and small-gap Ti~\cite{belzig_quasiclassical_1999}.

We obtain the interface parameter $Z$ from the ratio of the superconducting gap and the product of excess current, $I_\mathrm{exc}$, and normal state resistance extracted above $2\Delta_0$, $R_\mathrm{N}$, using Eq.~25 from~\cite{Niebler_2009}. For $I_\mathrm{exc}R_\mathrm{N}\approx209\,\unit{\micro\volt}$ at $V_\mathrm{g}=-1.5\,\unit{\volt}$, we estimate the interface parameter $Z\approx0.427$, which corresponds to transparent interfaces. This value of $Z$ motivates our choice of MAR positions in Fig.~\ref{fig:dc_measurements}(c). The multiple Andreev reflection is qualitatively explained by OBTK theory \cite{octavio1983subharmonic,flensberg1988subharmonic}, where it is found that the MAR features correspond to minima of the $dV/dI$ curve only for large interface parameter $Z\sim1$. For lower $Z\sim0.427$, as in our case, the MAR features correspond to the voltages above the minima, close to the maximum derivative of $R_\mathrm{diff}$ over $V$, as indicated in Fig.~\ref{fig:dc_measurements}(c).

We estimate the superconducting coherence length $\xi=\frac{{\hbar}v_\mathrm{F}}{2\Delta_\mathrm{0}}\approx1.94\,\unit{\micro\meter}$~\cite{PhysRevB.46.12573,Ohta1997} using the Fermi velocity $v_\mathrm{F}=8.3\times10^{5}\,\unit{\meter}/\unit{\second}$ from a $\mathbf{k{\cdot}p}$ calculation for our quantum well structure~\cite{beugeling2024kdotpymathbfkcdotmathbfptheorylattice}. By comparing this value to the device lengths, we conclude that all our devices are in the intermediate limit $\xi{\sim}l$, for which the size of the junction still has an effect on the ABS spectrum.

We investigate the temperature effect on short DC Josephson junction with side contacts in Fig.~\ref{fig:dc_measurements}(d). We see that the Multiple Andreev features $2\Delta$ and $\Delta$ ($2\Delta/3$ cannot be clearly viewed because of the plot limits) scale with temperature. We use the white dashed lines to show the result of an interpolation formula for the temperature dependent gap $\Delta(T)\approx\Delta_0\mathrm{tanh}(1.74 \sqrt{T_\mathrm{c}/T - 1})$ by plotting $\pm\Delta(T)$ and $\pm2\Delta(T)$. The extracted critical temperature $T_\mathrm{c}\approx945\,\unit{\milli\kelvin}$ is slightly different from the BCS prediction.

The smaller $R_\mathrm{diff}$ features in Fig.~\ref{fig:dc_measurements}(c,d) exhibit no temperature dependence and remain unaffected by the gate voltage. Therefore, they are likely unrelated to the intrinsic properties of the junction and may instead arise from the surrounding microwave environment, as previously discussed in~\cite{liu2024perioddoubling}. A junction in the finite-voltage state is susceptible to the environment resonances at frequencies matching its emission, which leads to the formation of voltage bumps in the time-averaged (DC) response.

Finally we note that the superconducting stack is slightly different for device D1-D2 and D3-D4, with more non-superconducting material present in D1-D2.

\section*{Appendix B: Tight-binding simulation}
\label{sec:appendix_tight_binding}

We simulate the ABS spectrum using the Bogoliubov-de Gennes formalism in a spinless tight-binding model on a square lattice with physical units, using similar approach as in~\cite{Ferrier2013,PhysRevResearch.3.L032009} with a discretized Hamiltonian:
\begin{equation}
	\mathcal{H} = \displaystyle\sum_{i}[E_i\sigma_z+\Delta_i \sigma_x]|i\rangle\langle i|-\displaystyle\sum_{i\neq j}t\delta_{i,j\pm1}\sigma_z|i\rangle\langle j|,
\end{equation}
where $\sigma_x$ and $\sigma_z$ are the Pauli matrices, $i,j$ represent the lattice indices, $t$ is the hopping energy, and $\delta_{i,j}$ is the Kronecker symbol. The superconducting gap, $\Delta_i$, is $\Delta_0$ in the superconducting electrodes and $0$ in the normal part. The on-site energy $E_{i}=E_\mathrm{site}-V_i$, where $E_\mathrm{site}$ is the same throughout the whole system and $V_i$ represents the local electrostatic potential. We take $V_i=0$ inside the superconductor and uniformly distributed $V_i\in[-E_\mathrm{dis},E_\mathrm{dis}]$ inside the normal area. We add a barrier potential $E_\mathrm{\delta}$ to the section in the middle of the junction, when applicable.

The phase difference across the JJ is set by applying a Peierls transformation inside the normal part of the system. To properly allow for a decay of the ABS wavefunctions with our large $\xi$ in the superconductor, the system has a total length of $10\,\unit{\micro\meter}$ with $500\,\unit{\nano\meter}$ long normal region in the middle (we verified that at this size of the leads stop influencing the ABS spectrum). The width of the system is chosen at $4\,\unit{\micro\meter}$. With a lattice size of $20\,\unit{\nano\meter}$ this results in a total system size of $500\times200$ sites.

The simulations are performed for $v_\mathrm{F}=8\times10^5\,\unit{\meter/s}$ and $k_\mathrm{F}=0.0393\,\unit{\nano\meter^{-1}}$. From this, we calculate the simulation parameters assuming a parabolic spectrum: $E_\mathrm{F}=10.35\,\unit{\milli\electronvolt}$, $t=16.75\,\unit{\milli\electronvolt}$ (before applying Peierls phase) and $E_\mathrm{site}=4t-E_\mathrm{F}=56.65\,\unit{\milli\electronvolt}$. We verify that $E_\mathrm{F}<t$ so that we operate in the parabolic part of the tight-binding spectrum and $\Delta_0{\ll}E_\mathrm{F}$ so that the Andreev approximation holds.

\begin{figure}
	\centering
	\includegraphics[scale=1.0]{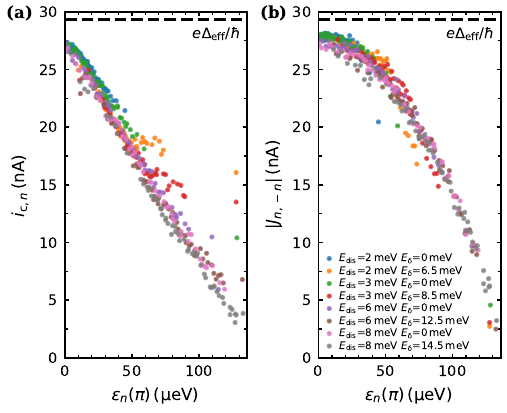}
	\caption{
		\textbf{Maximum supercurrent contribution and the non-diagonal elements of the current operator at phase $\pi$.}
		\textbf{(a)} Maximum supercurrent contribution $i_{\mathrm{c},n}=\mathrm{max}[i_n(\varphi)]$ as a function of the ABS energy at phase $\pi$, $\varepsilon_n(\pi)$, extracted from tight-binding simulations of a short JJ with varied disorder strength $E_\mathrm{dis}$ and potential barrier height $E_\mathrm{\delta}$ [symbols, see legend in panel (b)]. The dashed line indicates the supercurrent contribution ${e\Delta_\mathrm{eff}}/{\hbar}$, where the effective gap, $\Delta_\mathrm{eff}=\frac{2\xi}{2\xi+L}\Delta_0$.
		\textbf{(b)} Absolute value of the non-diagonal matrix element between electron-hole symmetric states at phase $\pi$, $|J_{n,-n}(\pi)|$, plotted against vs $\varepsilon_n(\pi)$, using the same simulation parameters as in (a).
	}
	\label{fig:matrix_element}
\end{figure}

We use the Kwant package~\cite{Groth_2014} to simplify the generation of the Hamiltonian matrix, which we solve for the eigenvalues and eigenstates using standard linear algebra methods. We find the matrix elements of the operator of current through a section at the superconducting leads in SI units using a Hermitian form that is valid for local currents with a Peierls phase:
\begin{equation}
\begin{aligned}
	J_{nm} & = \frac{je}{\hbar} \times \\
	\sum_{k} & t_u u_{n}^*(x_k,y_l) u_m(x_k,y_{l+1}) - t_u^* u_m(x_k,y_l)u_n^*(x_k,y_{l+1}) \\
	- & t_v v_n^*(x_k,y_l) v_m(x_k,y_{l+1}) + t_v^* v_m(x_k,y_l) v_n^*(x_k,y_{l+1}),
	\label{eqn:chitight}
\end{aligned}
\end{equation}
here $u(x,y)$ and $v(x,y)$ correspond to the components of the Bogoliubov wavefunction, $t_u$ and $t_v$ are the hopping parameters for corresponding wavefunction components, and $y_l$ is the section through which the current is calculated.

We checked the validity of our calculation by comparing the diagonal elements $\mathrm{Re}[J_\mathrm{nn}]$ and $i_n=\frac{2e}{\hbar}\frac{\partial\epsilon_n}{\partial\varphi}$ and find them matching closer than $1\%$.

We calculate the Josephson current using Eq.~\ref{eq:cpr_abs}, diagonal part of the microwave admittance using Eq.~\ref{eq:diagonal_admittance}, and non-diagonal part using Eq.~\ref{eq:kubo_losses}, properly taking into account the spin degeneracy.

To match the magnitude of the Josephson current and the normal state resistance, a Fermi wavevector $k_\mathrm{F}=0.0393\,\unit{\nano\meter^{-1}}$, is used for the simulations in Fig.~\ref{fig:tight_binging_spectra}. This value is approximately an order of magnitude smaller than $k_\mathrm{F}=\sqrt{{2\pi}n_e}\sim0.3\,\unit{\nano\meter^{-1}}$ calculated for the large carrier density of our devices. Simulations for $k_\mathrm{F}=0.3\,\unit{\nano\meter^{-1}}$ result in a much higher supercurrent ($I_\mathrm{c}$) magnitude. Increasing the disorder or barrier strength to match the $I_\mathrm{c}$ magnitude leads to a mismatch in the shape of the microwave admittance.
	
This discrepancy might be related to the interface between the superconductor and semiconductor, which can limit the number of states with a high interface transparency either due to interface disorder, or due to the fundamental effects of matching the electron wavefunctions in different materials~\cite{PhysRevB.59.10176}. While a semiconductor with higher carrier density is expected to host more ABS, the majority of them might be completely gapped, as we do not observe their contribution to the supercurrent, microwave loss at low temperatures, or to the normal charge transport. This could result of the interface effects, which produce different ABS transmission statistics compared to the disorder in the bulk of the semiconductor~\cite{RevModPhys.69.731}.

\section*{Appendix C: Matrix elements $J_{n,-n}$ and microwave loss at phase $\pi$}
\label{sec:appendix_matrix_elements}

\begin{figure}
	\centering
	\includegraphics[scale=1.0]{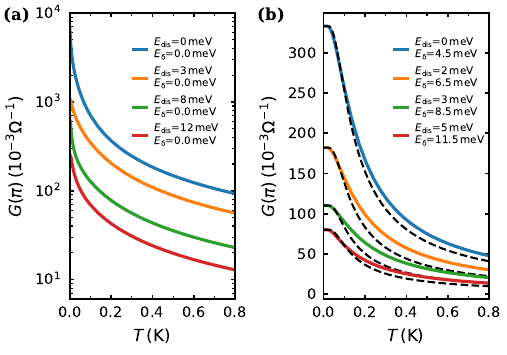}
	\caption{
		\textbf{Simulated temperature dependence of microwave loss at phase $\pi$.} The curves are generated for non-diagonal relaxation rate $\gamma_\mathrm{ND}=40\,\unit{\micro\electronvolt}$.
		\textbf{(a)} Temperature dependence for microwave loss at phase $\pi$, $G(\pi)$, generated for spectra without a gap at phase $\pi$, $E_\mathrm{g}(\pi)$. Legend lists the simulation parameters.
		\textbf{(b)} Same as (a), but for spectra with $E_\mathrm{g}\approx10\,\unit{\micro\electronvolt}$. The dashed lines are $G(\pi) \propto [1 - 2 f(E_\mathrm{eff},T)]$ dependencies with $E_\mathrm{eff}=17\,\unit{\micro\electronvolt}$.
	}
	\label{fig:g_pi_temperaure_dependence}
\end{figure}

At phase $\varphi=\pi$ the non-diagonal microwave loss in Eq.~\ref{eq:kubo_losses} is predominantly due to transitions between electron-hole symmetric states with opposite energies. This occurs because the matrix elements of the current operator between electron-hole symmetric states, $|J_{n,-n}(\pi)|$, dominate over all other off-diagonal matrix elements~\cite{Ferrier2013}.

\begin{figure*}
	\centering
	\includegraphics[scale=1.0]{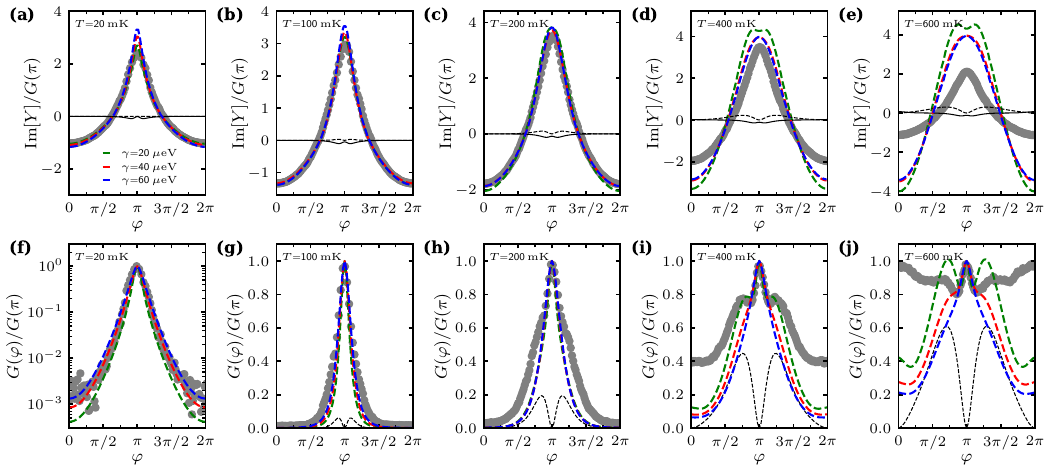}
	\caption{
		\textbf{Simulated microwave response and experimental data at higher temperatures.}
		\textbf{(a-e)} The imaginary part of microwave admittance $\mathrm{Im}[Y]$ at temperatures of $20\,\unit{\milli\kelvin}$, $100\,\unit{\milli\kelvin}$, $200\,\unit{\milli\kelvin}$, $400\,\unit{\milli\kelvin}$, and $600\,\unit{\milli\kelvin}$. The data are normalized by the microwave shunt conductance $G=\mathrm{Re}[Y]$ at phase $\pi$. Gray dots represent the measurements, while colored lines correspond to the simulations with different magnitude of the relaxation parameter $\gamma_\mathrm{ND}=\gamma_\mathrm{D}=\gamma$ [legend in panel (a)]. The dashed black line is the imaginary component of diagonal admittance $\mathrm{Im}[Y_\mathrm{D}]$, while the solid black line is the imaginary component of non-diagonal admittance $\mathrm{Im}[Y_\mathrm{ND}]$, both calculated for $\gamma=40\,\unit{\micro\electronvolt}$. For the simulation, we use a spectrum, generated with disorder magnitude $E_\mathrm{dis}=3\,\unit{\milli\electronvolt}$ and potential barrier $E_\mathrm{\delta}=8.5\,\unit{\milli\electronvolt}$.
		\textbf{(f-j)} The microwave shunt conductance $G=\mathrm{Re}[Y]$, corresponding to the data in (a-e), normalized by its value at phase $\pi$. The dashed black line represents the contribution of diagonal admittance $\mathrm{Re}[Y_\mathrm{D}]$ extracted from the numerical simulation for $\gamma=40\,\unit{\micro\electronvolt}$. In panel (a) $\mathrm{Re}[Y_\mathrm{D}]$ is too small and does not fit into the axis limits.
	}
	\label{fig:admittance_higher_t}
\end{figure*}

In~\cite{Ferrier2013}, it was observed that, for the lowest ABS, $|J_{1,-1}(\pi)|$ is nearly equal in magnitude to the maximum supercurrent carried by a state, $i_{\mathrm{c},1}$. Additionally, it was noticed that $|J_{n,-n}(\pi)|$ decreases slower than $i_{n}$ with increasing $n$. We verified this in our tight-binding simulations and extracted the trends of both $|J_{n,-n}(\pi)|$ and $i_{\mathrm{c},n}=\mathrm{max}(\mathrm{Re}[J_{nn}])$, as shown in Fig.~\ref{fig:matrix_element}. These results correspond to a Josephson junction with our device geometry. In the same figure, we present the data from simulations incorporating a varying amplitude of disorder, as well as cases with and without a gap opening at $\pi$.

For states with a vanishing gap, $\varepsilon_n(\pi){\to}0$, we observe that both the supercurrent contribution shown in Fig.~\ref{fig:matrix_element}(a) and $|J_{n,-n}(\pi)|$ in Fig.~\ref{fig:matrix_element}(b) saturate at approximately the same value. This value is slightly below the expected maximum supercurrent contribution, $i_\mathrm{max}=\frac{ev_\mathrm{F}}{L+2\xi}$~\cite{PhysRevB.46.12573}. By introducing an effective gap $\Delta_\mathrm{eff}=\frac{2\xi}{2\xi+L}\Delta_0$, which accounts for the effect of the junction length, this maximum supercurrent can also be expressed as $i_\mathrm{max}={e\Delta_\mathrm{eff}}/{\hbar}$.

As the gap $\varepsilon_n(\pi)$ increases, nearly all $i_{\mathrm{c},n}$ decrease linearly, with $i_{\mathrm{c},n}\to0$ as $\varepsilon_n(\pi)\to\Delta_0$. The values for $|J_{n,-n}(\pi)|$ decrease more in a parabolic manner, which initially drops slower, suggesting that the matrix elements of the lowest-lying ABS can be approximated as nearly constant for different $n$. This observation allows us to derive Eq.~\ref{eq:model_losses} and justifies the emergence of a characteristic transition energy, $E_\mathrm{eff}$, in the temperature dependence of the microwave loss.

Following the analysis of Eq.~\ref{eq:model_losses}, we simulate the temperature dependencies of the non-diagonal microwave loss at phase $\pi$, $G_\mathrm{ND}$, for junctions with and without gap at phase $\pi$. In Fig.~\ref{fig:g_pi_temperaure_dependence}(a) we show the results for simulations of ABS spectra without a gap at phase $\pi$. As expected from our analysis, these produce sharp dependencies, due to the dominating contribution of the states with $\epsilon_n(\pi)\to0$.

In contrast, the calculated dependencies for spectra with $E_\mathrm{g}(\pi)$ in Fig.~\ref{fig:g_pi_temperaure_dependence}(b) exhibit a dependence reminiscent of our experimental result, roughly following $G(\pi) \propto [1 - 2 f(E_\mathrm{eff},T)]$ with $E_\mathrm{eff}=17\,\unit{\micro\electronvolt}$.

\section*{Appendix D: Microwave admittance at higher temperature}
\label{sec:appendix_losses_high_t}

In this section, we compare the results of our tight-binding simulation with experimental data at higher temperatures and analyze the influence of diagonal and non-diagonal relaxation parameters on the produced result. In Fig.~\ref{fig:admittance_higher_t}(a-e), we show the normalized imaginary component of the microwave admittance $\mathrm{Im}[Y]$ for D1 at $V_\mathrm{g}=-2.5\,\unit{\volt}$ and temperatures: $20\,\unit{\milli\kelvin}$, $100\,\unit{\milli\kelvin}$, $200\,\unit{\milli\kelvin}$, $400\,\unit{\milli\kelvin}$, and $600\,\unit{\milli\kelvin}$ (gray symbols). The simulation output for various relaxation rates (we take $\gamma=\gamma_\mathrm{ND}=\gamma_\mathrm{D}$) is shown as dashed colored lines. For this analysis we take the ABS spectrum calculated with $E_\mathrm{dis}=3.0\,\unit{\milli\electronvolt}$ and $E_\mathrm{\delta}=8.5\,\unit{\milli\electronvolt}$, which resulted in best match at $T=20\,\unit{\milli\kelvin}$. Similarly, in Fig.~\ref{fig:admittance_higher_t}(c-j) we show the corresponding normalized microwave loss $G$.

From the data at $20\,\unit{\milli\kelvin}$ in Fig.~\ref{fig:admittance_higher_t}(a,f), we conclude that the simulations with a relaxation rate in the range $\gamma\in[20,40]\,\unit{\micro\electronvolt}$ result in a decent fit for our measured data. The results for a smaller value $\gamma=20\,\unit{\micro\electronvolt}$ yield a better fit to  $\mathrm{Im}[Y]$, whereas $\gamma=40\,\unit{\micro\electronvolt}$ provides a closer match for the microwave loss $G=\mathrm{Re}[Y]$.

The solid black line in Fig.~\ref{fig:admittance_higher_t}(a) represents $\mathrm{Im}[Y_\mathrm{ND}]$ for $\gamma=40\,\unit{\micro\electronvolt}$, while the dashed line is $\mathrm{Im}[Y_\mathrm{D}]$. Both contributions are negligible at $T=20\,\unit{\milli\kelvin}$, which confirms the applicability of the adiabatic limit and the validity of our method of CPR extraction. Subsequently, the change in $\mathrm{Im}[Y]/G(\pi)$ traces for different $\gamma$ happens exclusively due to the change of the normalization factor $G(\pi)$.

At higher temperatures, $T=100\,\unit{\milli\kelvin}$ [Fig.~\ref{fig:admittance_higher_t}(b)] and $T=200\,\unit{\milli\kelvin}$ [Fig.~\ref{fig:admittance_higher_t}(c)], the simulated and measured values of $\mathrm{Im}[Y]/G(\pi)$ remain in good agreement. The associated microwave loss, shown in Fig.~\ref{fig:admittance_higher_t}(g,h), is accurately reproduced around $\varphi=\pi$. However, the model does not capture the symmetric shoulders observed in the experimental data, which may result from the subtle features in the ABS spectrum. These shoulders bear a resemblance to the diagonal contribution $\mathrm{Re}[Y_\mathrm{D}]/G(\pi)$, illustrated by dashed lines for $\gamma=40\,\unit{\micro\electronvolt}$ in Fig.~\ref{fig:admittance_higher_t}(g,h). Adjusting the gap $E_\mathrm{g}(\pi)$ or the diagonal relaxation rate $\gamma_\mathrm{D}$ does not yield an improved fit to the experimental data.

At even higher, $T=400\,\unit{\milli\kelvin}$ [Fig.~\ref{fig:admittance_higher_t}(d)] and $T=600\,\unit{\milli\kelvin}$ [Fig.~\ref{fig:admittance_higher_t}(e)], the maximum experimental value of $\mathrm{Im}[Y]/G(\pi)$ begins to decrease, while the simulations continue to yield $\mathrm{Im}[Y]/G(\pi)\sim4$. This discrepancy arises from two well-understood effects. First, the superconducting gap $\Delta(T)$  in the aluminum electrodes diminishes at these temperatures, as confirmed by our DC measurements (\hyperref[sec:dc]{Appendix A}). As $\Delta(T)$ decreases, the supercurrent $I_\mathrm{J}$ is reduced, while the microwave loss $G$ increases. Second, as shown in Fig.~\ref{fig:losses}(i,j), our measurements reveal a larger increase in $G(0)$ than predicted by the model. This additional loss is attributed to continuum states in the superconducting loop, whose density of states can be accurately modeled only if the entire loop circumference is included. However, in our model, only approximately $\approx10\,\unit{\micro\meter}$ of the superconducting leads are considered due to computational limitations.


%\bibliography{manuscript.bib}
%\printfigures
\pagebreak
\clearpage
\widetext

\setcounter{equation}{0}
\setcounter{figure}{0}
\setcounter{table}{0}
\setcounter{page}{1}
\setcounter{section}{0}

\renewcommand{\thefigure}{S\arabic{figure}}
\renewcommand{\theequation}{S\arabic{equation}}
\renewcommand{\thesection}{NOTE \arabic{section}}
\renewcommand{\bibnumfmt}[1]{[S#1]}
\renewcommand{\citenumfont}[1]{S#1}

\begin{center}
	\textbf{\large Supplemental Material: Phase-dependent supercurrent and microwave dissipation of HgTe quantum well Josephson junctions}
\end{center}

\section{Sample fabrication}

\begin{figure*}[h]
	\centering
	\includegraphics[width=1.0\textwidth]{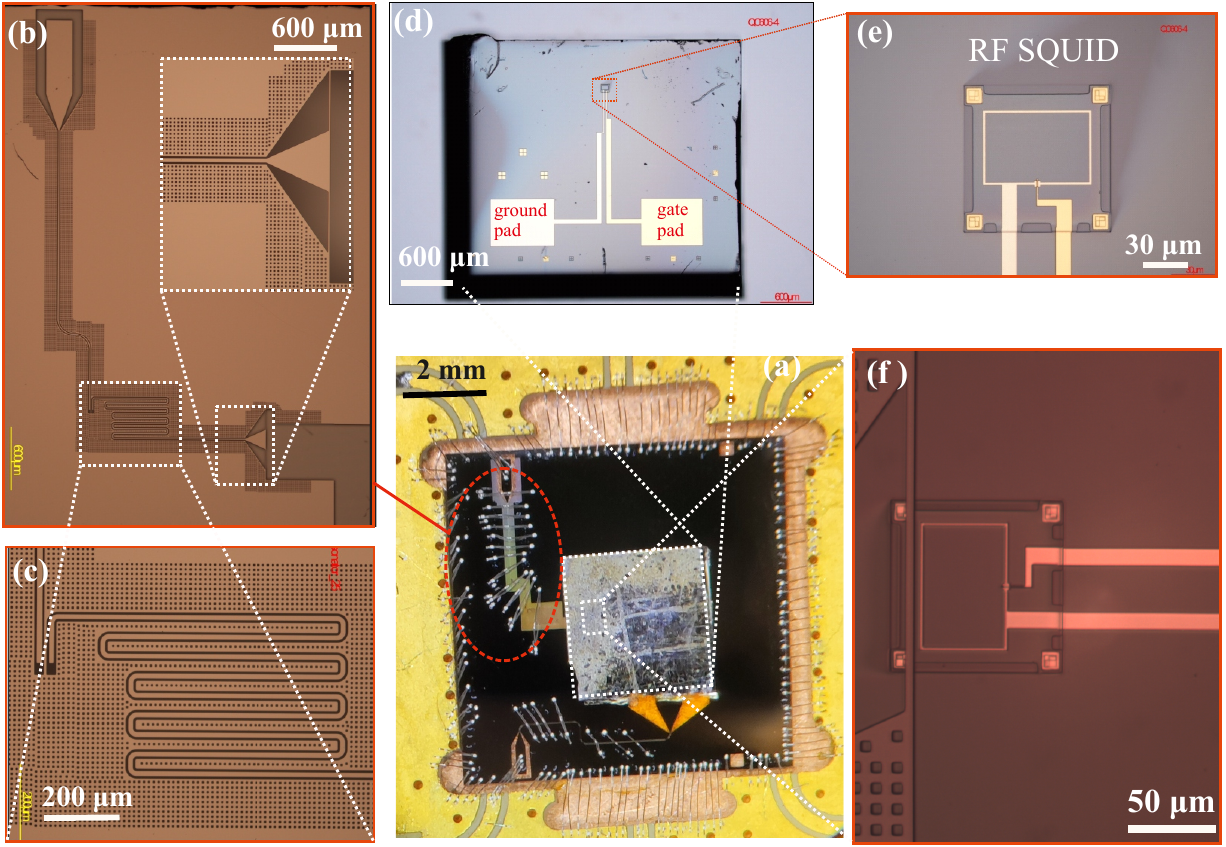}
	\caption{\textbf{Flip-chip device assembly.} \textbf{(a)} Image of a flip-chip device in the sample box. \textbf{(b)} Optical microscopy photograph of the niobium resonator with a zoomed-in view presented in \textbf{(c)}. \textbf{(d)} Optical microscopy photograph of the RF SQUID with a zoomed-in view presented in \textbf{(e)}. \textbf{(f)} Optical microscopy photograph of the RF SQUID aligned close to the grounded end of the resonator after the flip-chip process. RF squid loop is seen through the transparent sapphire along with Nb resonator structures on the back side of the substrate.}
	\label{sfig:flip_chip_devices}
\end{figure*}

A mesa structure is fabricated from an MBE-grown wafer containing $8\,\mathrm{nm}$ HgTe QW capped with a $\approx50\,\mathrm{nm}$ HgCdTe barrier by wet etching in a $\mathrm{KI:I_2:HBr}$ 1:4 $\mathrm{H_2O}$ solution. The HgTe QW material was characterized via Hall measurements on a macroscopic Hall bar on separate chip, cleaved from the same wafer. The Hall bar was fabricated without an electrostatic gate. Devices D1 and D2 were fabricated from wafer QC606 with carrier density $n_\mathrm{e}=1.77\times10^{11}\,\mathrm{cm^{-2}}$ and mobility $\mu=142\times10^{3}\,\mathrm{cm^{2}/(V{\cdot\,}s)}$. Devices D3 and D4 were fabricated from wafer Q3194 with low intrinsic carrier density, where characterization Hall measurements could only be performed after $2\,\unit{\second}$ LED illumination, resulting in a $n_\mathrm{e}=2.78\times10^{11}\,\mathrm{cm^{-2}}$ and $\mu=219\times10^{3}\,\mathrm{cm^{2}/(V{\cdot\,}s)}$. The estimate of bulk mean free path for D1 and D2 is: $l_\mathrm{mfp}=h\mu\sqrt{n_\mathrm{e}/2\pi}/e \approx 985\,\mathrm{nm}$. The mesa dimensions are confirmed by inspecting SEM micrographs of test structures.

Superconducting side contacts are deposited by e-gun evaporation of a $\mathrm{Ti/Al/Ti/Au}$ stack after in-situ Ar etching for $5\,\mathrm{s}$ (the exact deposited layer thicknesses as well as other parameters of the fabricated devices are listed in Table~\ref{tab:device_parameters}). The main $\mathrm{Al}$ layer was deposited in 4 cycles, each consisting of $20^\circ$, $-20^\circ$, $0^\circ$ evaporation angle steps, while the other layers in a single cycle with two $\pm20^\circ$ steps.

The gate structure is fabricated by first growing a $\mathrm{HfO_2}$ dielectric layer by atomic layer deposition, followed by e-beam evaporation of a $5\,\unit{\nano\meter}$ $\mathrm{Ti}$ sticking layer and a $150\,\unit{\nano\meter}$ $\mathrm{Au}$ film at a $0^\circ$ angle.

In small JJ devices, we detect a carrier density much higher than in a macroscopic Hall bar fabricated from the same material. In the $200\,\unit{\nano\meter}$ long DC device D4, we observe this directly via a measurement of magnetoresistance oscillations (see Note 4). We identified two sources of the carrier density increase in our side-contacted JJs. The $5\,\mathrm{s}$ in-situ Ar etching seems to be main cause, resulting in unintentional doping of the near-contact area, leading to an increase by up to $\sim10^{12}/\unit{\centi\meter}^{2}$ for the smallest mesas. Without this step in the sample processing, devices with virtually no $n_\mathrm{e}$ increase can be fabricated. This does require additional measures to prevent interface oxidation. Additionally, gate fabrication increases $n_\mathrm{e}$ by $\approx1\times10^{11}/\unit{\centi\meter}^{2}$ for small devices~\cite{S_Liang_2024}.

\begin{table}
	\caption{\label{tab:device_parameters}Parameters of the devices studied.}
	\begin{ruledtabular}
		\begin{tabular}{llllll}
			Device & Contact geometry & JJ Length & JJ Width & Wafer & Superconducting contact stack \\
			D1 & RF SQUID $70\,\unit{\micro\meter}\times47\,\unit{\micro\meter}$ & $500\,\unit{\nano\meter}$ & $3.8\,\unit{\micro\meter}$ & QC606 & $9\,\unit{\nano\meter}\,\mathrm{Ti}$, $115\,\unit{\nano\meter}\,\mathrm{Al}$, $7\,\unit{\nano\meter}\,\mathrm{Ti}$, $15\,\unit{\nano\meter}\,\mathrm{Au}$\\
			D2 & RF SQUID $70\,\unit{\micro\meter}\times47\,\unit{\micro\meter}$ & $500\,\unit{\nano\meter}$ & $3.8\,\unit{\micro\meter}$ & QC606 & $9\,\unit{\nano\meter}\,\mathrm{Ti}$, $115\,\unit{\nano\meter}\,\mathrm{Al}$, $7\,\unit{\nano\meter}\,\mathrm{Ti}$, $15\,\unit{\nano\meter}\,\mathrm{Au}$\\
			D3 & RF SQUID $40\,\unit{\micro\meter}\times20\,\unit{\micro\meter}$ & $200\,\unit{\nano\meter}$ & $3.8\,\unit{\micro\meter}$ & Q3194 & $5\,\unit{\nano\meter}\,\mathrm{Ti}$, $100\,\unit{\nano\meter}\,\mathrm{Al}$, $5\,\unit{\nano\meter}\,\mathrm{Ti}$, $10\,\unit{\nano\meter}\,\mathrm{Au}$\\
			D4 & 4 terminal DC & $200\,\unit{\nano\meter}$ & $3.8\,\unit{\micro\meter}$ & Q3194 & $5\,\unit{\nano\meter}\,\mathrm{Ti}$, $100\,\unit{\nano\meter}\,\mathrm{Al}$, $5\,\unit{\nano\meter}\,\mathrm{Ti}$, $10\,\unit{\nano\meter}\,\mathrm{Au}$\\
		\end{tabular}
	\end{ruledtabular}
\end{table}

The RF SQUID loops are of rectangular shape with a trace width of $1.5\,\unit{\micro\meter}$. The RF SQUID loop and gate are connected to large pads [Fig.~\ref{sfig:flip_chip_devices}(d)] that are used for making electrical and mechanical connections to the resonator chip.

A quarter-wavelength superconducting microstrip line resonator, with one end capacitively coupled to the transmission line and the other directly connected to the ground [meander line on Fig.~\ref{sfig:flip_chip_devices}(b) and (c)], is fabricated separately by first sputtering $150\,\mathrm{nm}$ $\mathrm{Nb}$ onto a pre-cleaned $0.5\,\mathrm{mm}$ thick sapphire substrate, followed by etching the pattern through $\mathrm{Al}$ mask with inductively coupled plasma. Finally the $\mathrm{Al}$ mask is removed by selective chemical etching and $\mathrm{Ti/Au}$ layers are placed onto the pads, designed for contacting the HgTe chip.

\section{Flip-chip technique}

In this section, we discuss how the HgTe sample is glued to the resonator to form a flip-chip device. To achieve a large inductive coupling between the resonator and the RF SQUID loop, the loop should be placed close to the grounded end of the resonator. The precision alignment of RF SQUID loop with the resonator is achieved by using a SUSS MJB 3 mask aligner with visual control of the loop position through the transparent sapphire substrate of the resonator [Fig.\,\ref{sfig:flip_chip_devices}(f)]. For that, the sapphire resonator is mounted on a home-made chip holder with a view window. The RF SQUID chip is placed on the lower chip holder located right below the resonator. By adjusting the position of the sapphire chip, we can align the superconducting loop as desired [Fig.~\ref{sfig:flip_chip_devices}(f)]. The electrical and mechanical contacts between chips are made via a small amount of conductive silver epoxy ("Master Bond EP21TDCS-LO") placed on the gate and ground contact pads [Fig.~\ref{sfig:flip_chip_devices}(d)]. After aligning and connecting the chips, they are left in the mask aligner at room temperature for $\sim 24$ hours to let the epoxy cure.

There are several advantages in using the flip-chip technique. Obtaining a larger internal quality factor $Q_\mathrm{i}$ results in higher sensitivity of the resonator as a detector. Sapphire substrates have considerably less losses than CdTe and resonators with quality factors in excess of $Q_\mathrm{i}\sim 10^{5}$ are routinely fabricated. This value is two orders of magnitude larger compared to resonators we fabricated on CdHgTe. Moreover, by fabricating the resonator structure on a separate substrate we reduce the number of lithography steps to which our HgTe layer is exposed, thus preserving the quality of the material.

Figure~\ref{sfig:flip_chip_devices}(a) shows the flip-chip device inside the sample box. The electrodes on the chip are wire-bounded with $\mathrm{Al}$ bonds to the $50\,\unit{\ohm}$ co-planar waveguide of a microwave chip carrier PCB. To avoid spurious resonances, we use multiple $\mathrm{Al}$ bonds to connect the separated parts of the resonator ground plane together and with the PCB ground plane.

\newpage
\section{Measurement setup}

\begin{figure*}[h]
	\centering
	\includegraphics[width=0.8\textwidth]{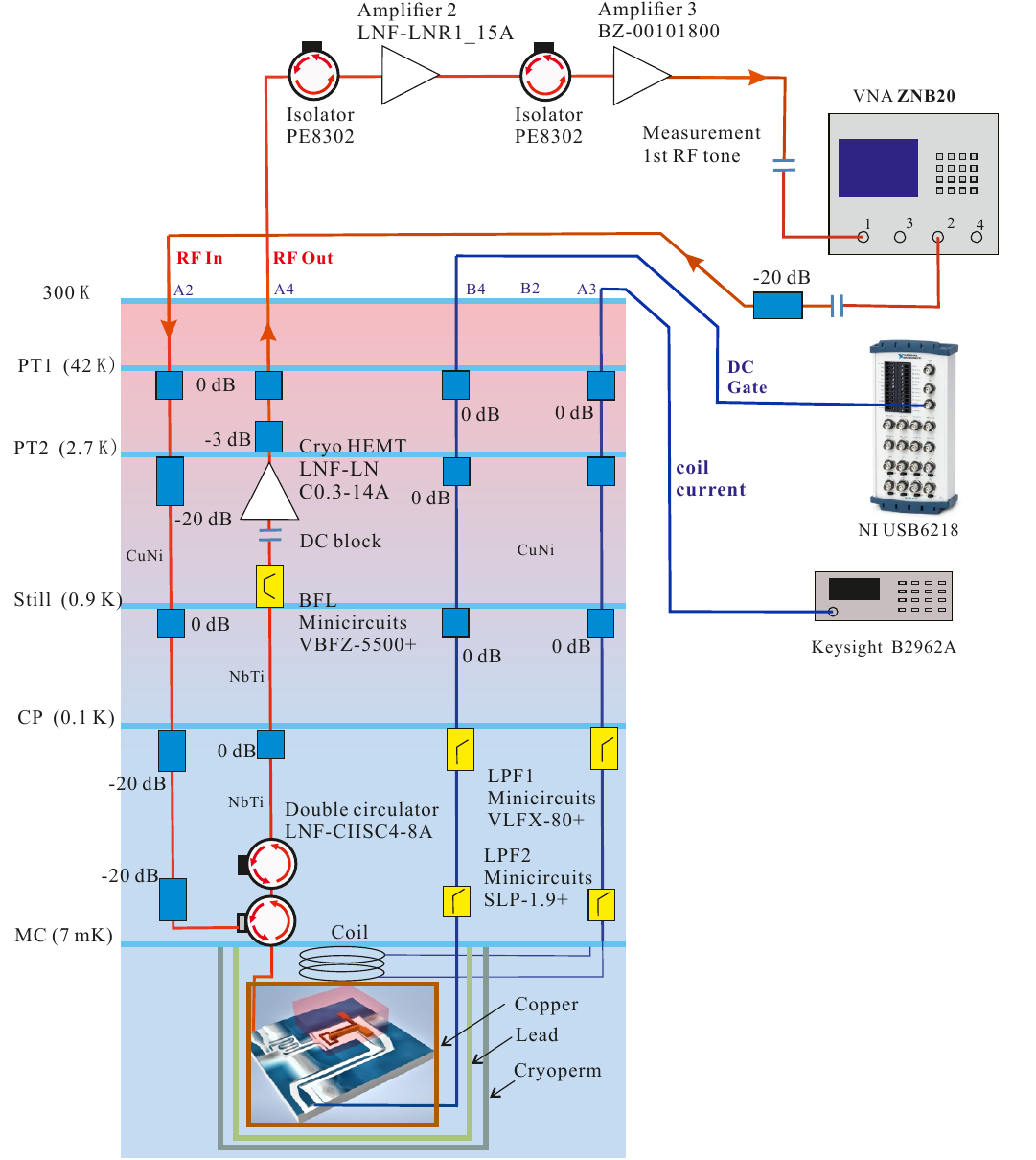}
	\caption{\textbf{Wiring diagram of experimental setup.} Devices are cooled down in a dilution fridge and are thoroughly shielded from the ambient magnetic field with cryogenic grade mu-metal (Cryoperm) and superconducting $\mathrm{Pb}$ screens. The microwave measurement wiring is indicated by the red lines. Large attenuation of the thermal noise combined with cryogenic and room-temperature amplification allows us to measure the microwave admittance in the linear response regime.}
	\label{sfig:measurement_setup}
\end{figure*}

Fig.~\ref{sfig:measurement_setup} shows the detailed schematic measurement setup. The flip-chip device is glued into a copper box and cooled down in a BlueFors LD400 dilution fridge with base temperature $\approx7\,\unit{\milli\kelvin}$. A home-made superconducting coil, sitting on top of the copper sample box, allows to control the flux through the RF SQUID loop. The coil current is sourced by a Keysight B2962A sourcemeter unit in current mode and supplied via a low-pass-filtered RF line. Two layers of magnetic shielding are used to reduce flux fluctuations as well as to minimize the number of vortices trapped in the Nb resonator due to the ambient magnetic field. Shielding is provided by a small, home made lead box and an outer Cryoperm can. A coaxial line connects to the SMP connector soldered on a microwave PCB.

In the microwave input line, several attenuators are placed at different cooling stages of the refrigerator to prevent external thermal noise from reaching the sample. The total attenuation (including losses in the cables) is around $-90$ dB. The $0$ dB attenuators thermally anchor the inner conductors of the coax lines without signal attenuation. The microwave signal reaching the resonator is less than $-110$ dBm in order to work in the linear response regime (see Fig.~\ref{sfig:linear_regime}). Two $4-8\,\unit{\giga\hertz}$ circulators (LNF CIISC4-8A) are placed at the mixing chamber plate to prevent noise from the amplifier reaching the sample. 

The microwave signal, injected at the lower circulator, probes the resonator response around the resonance frequency $f_\mathrm{r}$. The reflected signal from the resonator goes through the two aforementioned circulators into a cryogenic HEMT amplifier (LNF-LNC0.3-14A) with $37\,\unit{\decibel}$ gain, placed at the $2.7\,\unit{\kelvin}$ plate. To minimize losses in the reflected signal, superconducting coaxial cables (Coax-Co SC-086-50-NbTi-NbTi) are used between the circulator and the cryogenic amplifier. The output line from the cryogenic amplifier to room temperature is a CuNi coax (CoaX-Co SC-086-50-CN-CN). The reflected signal is further amplified with two additional room temperature amplifiers (LNF-LNR1-15A with $37\,\unit{\decibel}$ gain and BZ-00101800 with $32\,\unit{\decibel}$ gain). The reflection coefficient is directly measured using a vector network analyzer (Rohde \& Schwarz ZNB20). DC gate voltage is applied through a low-pass-filtered coaxial line using a NI USB6218 digital-to-analog converter.

\section{Carrier density extraction for DC device D4}

\begin{figure*}[h]
	\centering
	\includegraphics[scale=1]{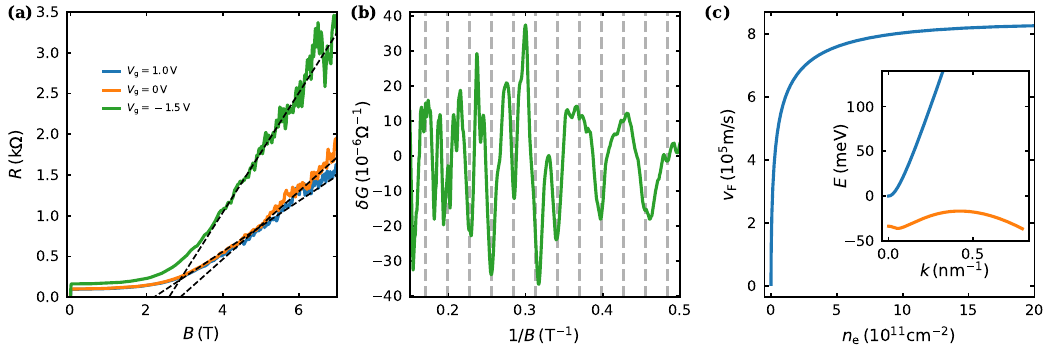}
	\caption{
		\textbf{Magnetoresistance measurements of DC device D4.}
		\textbf{(a)} Magnetic field ($B$) dependence of two-terminal linear response resistance, $R$, for device D4, measured at gate voltages $V_\mathrm{g}=1.0\,\unit{\volt}$, $V_\mathrm{g}=0\,\unit{\volt}$, and $V_\mathrm{g}=-1.5\,\unit{\volt}$. The dashed black lines represent linear slopes, corresponding to $n_\mathrm{e}=20\times10^{11}\,\unit{\centi\meter^{-2}}$, $n_\mathrm{e}=15\times10^{11}\,\unit{\centi\meter^{-2}}$, and $n_\mathrm{e}=8.5\times10^{11}\,\unit{\centi\meter^{-2}}$.
		\textbf{(b)} Oscillating part of conductance ${\delta}G$, after removing the monotonic background from $G=1/R$ for $V_\mathrm{g}=-1.5\,\unit{\volt}$. Dashed lines represent the positions of integer filling factors for $n_\mathrm{e}=8.5\times10^{11}\,\unit{\centi\meter^{-2}}$.
		\textbf{(c)} Fermi velocity as a function of carrier density in the bulk conduction band, extracted from the $\mathbf{k{\cdot}p}$ calculation for $8\,\unit{\nano\meter}$ quantum well. The inset shows calculated bulk spectrum close to the $\Gamma$ point with conduction (blue) and valence (orange) bands.
	}
	\label{sfig:hall}
\end{figure*}

We extract the carrier density of the side-contacted JJ device D4 by measuring the magnetic field dependence of a linear response resistance $R$ in Fig.~\ref{sfig:hall}(a). The superconductivity of the JJ electrodes is suppressed at small magnetic field and does not influence the measurements above that. The measured two-terminal resistance is a combination longitudinal and Hall resistances. At high $B$ and for large aspect ratio (20) the Hall component dominates. This allows for two methods of carrier density ($n_\mathrm{e}$) extraction, which yield identical results.

First, the carrier density can be extracted by fitting the slope at larger field of traces [dashed lines in Fig.~\ref{sfig:hall}(a)]. We fit the traces above $4\,\unit{\tesla}$ with $R=R_\mathrm{off} +  B / n_\mathrm{e} / e$, which produces $n_\mathrm{e}=20\times10^{11}\,\unit{\centi\meter^{-2}}$, $n_\mathrm{e}=15\times10^{11}\,\unit{\centi\meter^{-2}}$, and $n_\mathrm{e}=8.5\times10^{11}\,\unit{\centi\meter^{-2}}$ for $V_\mathrm{g}=1.0\,\unit{\volt}$, $V_\mathrm{g}=0\,\unit{\volt}$, and $V_\mathrm{g}=-1.5\,\unit{\volt}$ respectively.

Second, we analyze the resistance oscillations, arising from the formation of Landau levels in the electron system of the semiconductor. To do so, we convert the data in Fig.~\ref{sfig:hall}(a) to conductance $G=1/R$, remove the non-oscillating background by subtracting an 8th degree polynomial fit of the whole data range, and smooth with a moving average filter. The resulting oscillatory part ${\delta}G$ for $V_\mathrm{g}=-1.5\,\unit{\volt}$ is shown in Fig.~\ref{sfig:hall}(b) as a function of inverse magnetic field, $B^{-1}$. We plot the positions corresponding to the integer filling factors, $\nu$, ${B_\nu}^{-1}=e{\nu}/(n_\mathrm{e}h)$ for $n_\mathrm{e}=8.5\times10^{11}\,\unit{\centi\meter^{-2}}$ in Fig.~\ref{sfig:hall}(b) as gray dashed lines. At large field ($B^{-1}<0.35\,\unit{\tesla^{-1}}$) integer filling factors align with the minima of ${\delta}G$. At smaller fields ($B^{-1}>0.35\,\unit{\tesla^{-1}}$), oscillations with twice larger period are observed, which corresponds to incomplete spin splitting of the Landau levels.

The same two methods can be applied to the samples with lower carrier density~\cite{S_liu2024perioddoubling}, where the Hall resistance grows faster with field and Quantum Hall plateaus are observed.

Based on the extracted values of $n_\mathrm{e}$, other properties of the 2D electron gas can be determined if the band structure of the material is known. For this purpose, we carry out a band structure calculation using the $\mathbf{k{\cdot}p}$ method as implemented by the \textbf{kdotpy} software in Ref.~\cite{S_beugeling2024kdotpymathbfkcdotmathbfptheorylattice}. In Fig.~\ref{sfig:hall}(c), we present the results, obtained from a calculation with command line:

\texttt{
	kdotpy-2d.py 8o k 0 1.0 / 1000 kphi 45 msubst CdZnTe 4\% mlayer HgCdTe 68\% HgTe HgCdTe 68\% llayer 10 8.0 10 zres 0.25 targetenergy 0 neig 100 erange -60 200 split 0.01 legend char plotstyle auto dos banddos noax cardens 0.006 obs
}

This model considers an infinite 2DEG layer so that only the bulk spectrum is calculated. The main axes in Fig.~\ref{sfig:hall}(c) show the Fermi velocity, $v_\mathrm{F}$, calculated as a derivative of spectrum as a function of $n_\mathrm{e}$. The inset shows the band structures of the highest valence band and lowest conduction band subbands close to the $\Gamma$ point as a function of wavevector $k$. The Fermi wavevector is calculated from the carrier density: $k_\mathrm{F}=\sqrt{{2\pi}n_e}$.

\section{Raw data of the microwave measurements}

\begin{figure*}[h]
	\centering
	\includegraphics[scale=1]{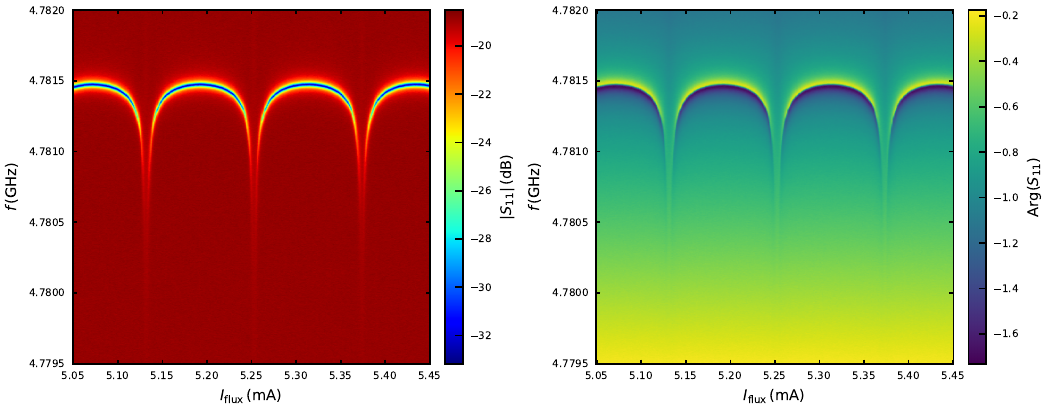}
	\caption{\textbf{Microwave measurements of the reflected signal.} Left (right) plot shows the absolute value (phase) of the microwave signal reflected by the resonator.  $I_\mathrm{flux}$ is the current applied on the home-made coil to control the flux in the loop.}
	\label{sfig:mag-arg}
\end{figure*}

\begin{figure*}[h]
	\centering
	\includegraphics[scale=0.92]{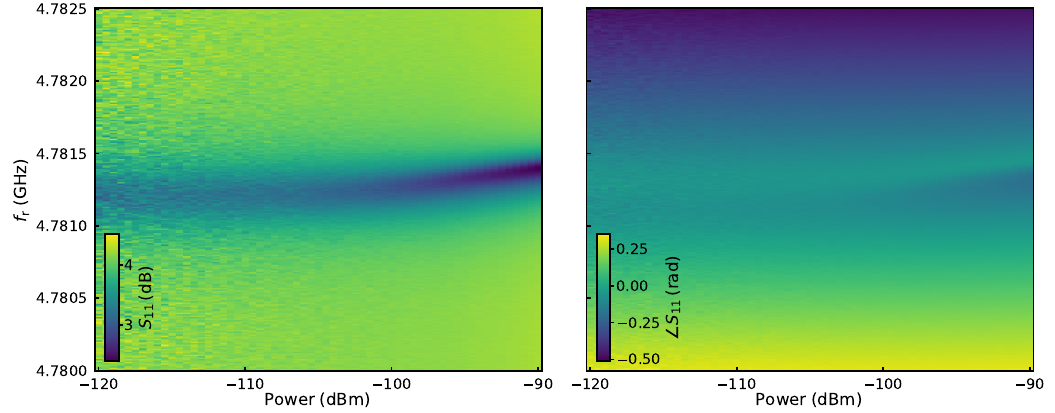}
	\caption{\textbf{Power dependence of the microwave response.} The magnitude (left) and phase (right) of the microwave response as a function of microwave power for the device D1 at the lowest temperature and $V_\mathrm{g}=-2\,\unit{\volt}$, biased close to phase bias $\varphi=\pi$. We avoid non-linearity in the experiment by using sufficiently low excitation ($-110\,\mathrm{dBm}$).}
	\label{sfig:linear_regime}
\end{figure*}

The microwave admittance measurement is performed by detecting the reflected signal from the superconducting resonator. By supplying small external magnetic field, we are able to tune the phase difference of the junction. For the microwave response, only the region adjacent to the resonance frequency is of interest. We record both magnitude and phase of the reflected signal.

In Fig.~\ref{sfig:mag-arg}, we show the absolute value and phase of the reflected signal as a function of magnet current, $I_{flux}$, and frequency. The data is taken by sweeping the frequency for fixed values of the magnetic field (junction phase). This procedure is carried out for all admittance measurements. The resonance is observed as a dip in reflected signal, accompanied by a rotation of the microwave phase. We clearly see three periods of the resonance frequency in both components. Phase $\pi$ is identified with a large negative shift in the resonance frequency, as the Josephson admittance $(j{\omega}L_\mathrm{J})^{-1}$ reaches its maximum.

In Fig.~\ref{sfig:linear_regime} we show the dependence of the microwave response (magnitude and phase) on the applied readout power. The data is measured for device D1 close to phase bias $\varphi=\pi$. At power level above $-105\,\unit{\decibel}\mathrm{m}$ a notable shift of the resonance frequency is observed, which obviously should be avoided. We use a power level of $-110\,\unit{\decibel}\mathrm{m}$ for our measurements to achieve that and to have a large enough signal-to-noise ratio.

\section{Fitting the resonance of the reflection curves}

\begin{figure*}[h]
	\centering
	\includegraphics[scale=1]{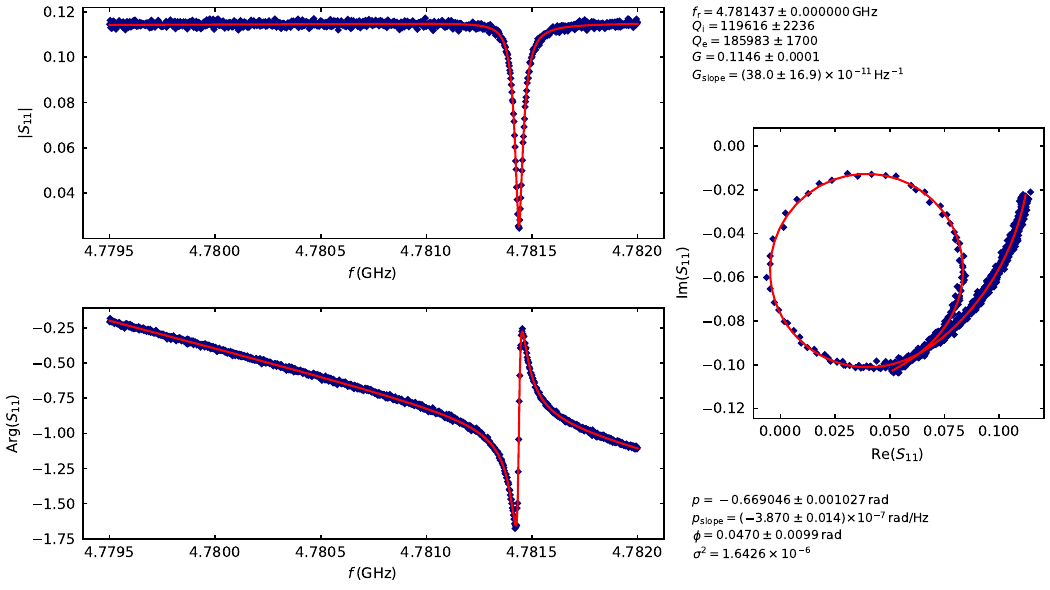}
	\caption{\textbf{Example of resonance parameter extraction via a direct fit of real and imaginary parts of reflected microwave signal $S_{11}$.} Left top and bottom panels show the absolute value and phase of the reflected signal, respectively, (blue points) together with the fit curves obtained using the model (red lines). The right panel shows the same data in I-Q space. The insets gives the fit parameters.}
	\label{sfig:fit_example}
\end{figure*}

To analyze the reflected signal, three relevant parameters need to be extracted for each frequency trace: the resonance frequency $f_\mathrm{r}$, the internal quality factor $Q_\mathrm{i}$ and the external quality factor $Q_\mathrm{e}$ . Both $f_\mathrm{r}$ and $Q_\mathrm{i}$ depend strongly on the properties of the RF SQUID and its inductive coupling to the resonator. $Q_\mathrm{e}$, on the other hand, is primarily determined by the coupling capacitor on the resonator chip and should be constant.

Properly extracting the parameters of the resonator response requires a simultaneous fit of both components (magnitude/phase or real/imaginary part), while the output signal is affected by measurement noise, phase delay and attenuation. We do this by a direct least-square fit on the signal components~\cite{S_Haller2022}.

According to the model, the observed complex reflected signal is described by~\cite{S_Haller2022}:
\begin{equation}
	\begin{split}
		S_{11} = (G + G_\mathrm{slope}(f - f_0))e^{j(p + p_\mathrm{slope}(f - f_0))} \times 
		\left[1 - \frac{Q_\mathrm{e}^{-1}}{(Q_\mathrm{i}^{-1}+Q_\mathrm{e}^{-1})/2 + jdf}e^{j\phi}\right]
	\end{split}
\end{equation}
Here $df = (f-f_\mathrm{r})/f_\mathrm{r}$ is the detuning from the resonance frequency $f_\mathrm{r}$. $G$ and $p$ are the gain and phase delay at a chosen reference frequency $f_0$. $G_\mathrm{slope}$ and $p_\mathrm{slope}$ are the corresponding frequency slopes of these parameters. We use the same $f_0$ for analysis within a data set, so that the stability of other parameters can be assessed. $\phi$ represents the phase rotation of absorbed signal.

We estimate the variance of the extracted parameters using the Hessian $\nabla{f(x)}\approx{\mathbf{J}(x)^\mathrm{T}\mathbf{J}(x)}$ approximation through the Jacobian $\mathbf{J}(x)$, thus employing the same approach as Ref.~\cite{S_Probst2015}. The statistical errors are estimated at the $95\%$ percentile (Fig.~\ref{sfig:fit_example}).

The parameters $p_\mathrm{slope}$, $G_\mathrm{slope}$ and $\phi$ should remain constant throughout the data set as we vary the junction-related parameters ($V_\mathrm{g}$, $\varphi$) since these are primarily related to resonator and transmission line properties. Additionally, due to the small statistical significance, the estimates based on a single curve have large relative error. To further improve the quality of our fits, we first fit the whole data set keeping these parameters free, later constraining them to the average of estimates from multiple frequency traces, resulting in a computationally-effective alternative to a simultaneous fit of the whole data set.

\section{Self-consistent extraction of the current-phase relation}

\begin{figure*}[h]
	\centering
	\includegraphics[scale=1.0]{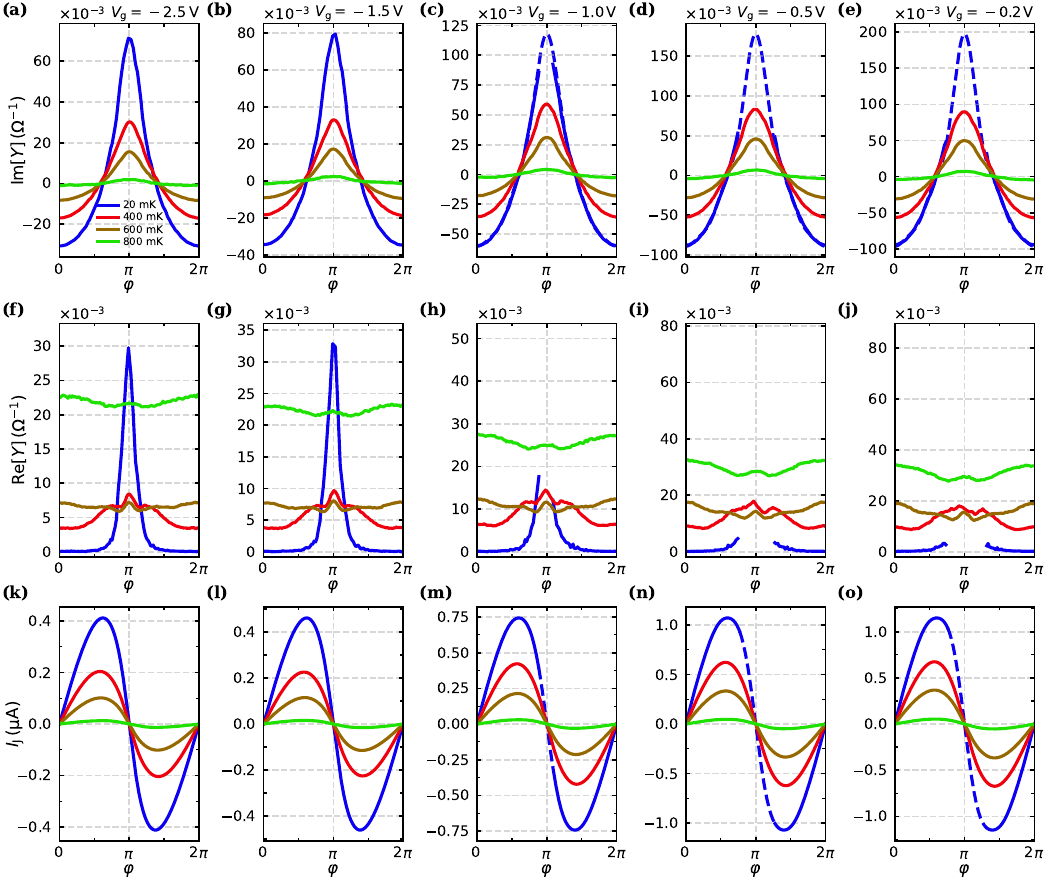}
	\caption{\textbf{Components of the microwave admittance $Y$ for device D1.} \textbf{(a-e)} Imaginary component $\mathrm{Im}[Y]$ as a function of JJ phase $\varphi$, primarily related to the Josephson inductance $L_\mathrm{J}^{-1} = \frac{2\pi}{\Phi_0}\frac{\partial I_\mathrm{J}}{\partial \varphi}$ for several temperatures [legend in panel (a)]. The dashed lines in some $T=20\,\unit{\milli\kelvin}$ traces mark the region where the admittance is inaccessible and is obtained from a fit assuming a CPR composed of the five lowest sine harmonics. \textbf{(f-j)} Real part of microwave response, $\mathrm{Re}[Y]$ labeled throughout the manuscript as the microwave shunt conductance $G$. \textbf{(k-o)} Current-phase relation (CPR), extracted from $\mathrm{Im}[Y]$ in (a-e) using the self-consistent extraction procedure.
	}
	\label{sfig:admittance_d1}
\end{figure*}

Following~\cite{S_Haller2022}, we have established a procedure connecting the offset of the measured resonance frequency and the change of the resonator quality factor to the parameters of the JJ. For convenience, in this section we represent the measured RF SQUID admittance $Y_\mathrm{m}=Z_\mathrm{m}^{-1}$ through the measured inductance $L_\mathrm{m}$ and conductance $G_\mathrm{m}$ as $Y_\mathrm{m} = (j{\omega}L_\mathrm{m})^{-1} + G_\mathrm{m}$.

The measured dissipationless admittance component is then proportional to the frequency shift~\cite{S_Haller2022}:
\begin{equation}
	\mathrm{Im}[Y_\mathrm{m}] = -({\omega}L_\mathrm{m})^{-1} = -\frac{{\pi}Z_\mathrm{r}}{2({\omega} M)^2} \frac{f_\mathrm{r} - f_\mathrm{r,0}}{f_\mathrm{r,0}}= -\frac{\pi^2}{8}\frac{L_\mathrm{r}}{{\omega}M^2}\frac{f_\mathrm{r} - f_\mathrm{r,0}}{f_\mathrm{r,0}},\label{seq:f_res}
\end{equation}
where $Z_\mathrm{r}\approx50\,\unit{\Omega}$ is the characteristic impedance of the resonator and $L_\mathrm{r}\approx2.13\,\unit{\nano\henry}$ is the full inductance of the resonator, while $f_\mathrm{r,0}$ is the resonance frequency of the unloaded resonator. 

Similarly, the internal quality factor can be converted to the dissipation-related component of the admittance using the unloaded resonator quality factor $Q_\mathrm{i,0}$:
\begin{equation}
	\mathrm{Re}[Y_\mathrm{m}] = G_\mathrm{m} = \frac{{\pi}Z_\mathrm{r}}{4({\omega}M)^2}(Q_\mathrm{i}^{-1} - Q_\mathrm{i,0}^{-1}) = \frac{\pi^2}{16}\frac{L_\mathrm{r}}{{\omega}M^2}(Q_\mathrm{i}^{-1} - Q_\mathrm{i,0}^{-1}).\label{seq:q_i}
\end{equation}

The measured RF SQUID admittance is different from the JJ admittance $Y=Z^{-1}$ due to the contribution of the loop inductance $L_\mathrm{loop}$. To account for it, we have to sum the corresponding impedances, so that $Z_\mathrm{m} = Z + j{\omega}L_\mathrm{loop}$. This allows us to express the measured admittance components by JJ admittance components $Y=(j{\omega}L_\mathrm{J})^{-1} + G$ and $L_\mathrm{loop}$. We expand the measured admittance:
\begin{equation}
	Y_\mathrm{m} = (Z + j{\omega}L_\mathrm{loop})^{-1}
	= \frac{Y}{1 + j{\omega}L_\mathrm{loop}Y}
	= \frac{(j{\omega}L_\mathrm{J})^{-1}+G}{1 + j{\omega}L_\mathrm{loop}[(j{\omega}L_\mathrm{J})^{-1}+G]}
	\label{seq:chi_screening}
\end{equation}
and split the real and imaginary parts, obtaining:
\begin{equation}
	L_\mathrm{m}^{-1} = \frac{L_\mathrm{J}^{-1}(1 + L_\mathrm{loop}/ L_\mathrm{J}) + {\omega}^2L_\mathrm{loop}G^{2}}{(1 + L_\mathrm{loop}/ L_\mathrm{J})^2 + ({\omega}L_\mathrm{loop}G)^2},\,
	G_\mathrm{m} = \frac{G}{(1 + L_\mathrm{loop}/ L_\mathrm{J})^2 + ({\omega}L_\mathrm{loop}G)^2}\label{seq:jj_params}
\end{equation}

We consider the adiabatic limit, hence the JJ microwave inductance is dominated by the Josephson contribution:
\begin{equation}
	L^{-1}_\mathrm{J}=\frac{2\pi}{\Phi_0}\frac{{\partial}I_\mathrm{J}(\varphi)}{{\partial}\varphi}.
	\label{seq:josephson_inductance}
\end{equation}
Additionally one has to consider the partial screening of the external flux $\varphi_\mathrm{ext}$ by the supercurrent $I_\mathrm{J}$. The actual phase bias of the junction becomes:
\begin{equation}
	\varphi = \varphi_\mathrm{ext} - \frac{2\pi}{\Phi_0}L_\mathrm{loop}I_\mathrm{J}(\varphi).\label{seq:flux_screening}
\end{equation}
To obtain $I_\mathrm{J}(\varphi)$ and $G(\varphi)$ from the measured $f_{\mathrm{r}}$ and $Q_{\mathrm{i}}$ one has to find a self-consistent solution for $I_\mathrm{J}(\varphi)$. We use an iterative approach, introduced in Ref.~\cite{S_Haller2022}, but modify it to also be valid when the condition ${\omega}L_\mathrm{loop}G{\ll}1$ does not hold, as described below.

We start by approximating the current-phase relation as Fourier series of sine harmonics, which is valid under preserved time-reversal symmetry:
\begin{equation}
	I_\mathrm{J}(\varphi) = \sum_{k\ge1} A_k\sin(k\varphi).\label{seq:cpr_ansatz}
\end{equation}
Initially, we set $G_0=0$ and phase $\varphi_0=\varphi_\mathrm{ext}$. Each iteration $i$ consists of the following steps:
\begin{enumerate}
	\item A least-squares fit of $f_\mathrm{r}$ as a function of $\varphi=\varphi_i$ using Eqs.~\ref{seq:f_res}, \ref{seq:jj_params}, \ref{seq:cpr_ansatz}, and $G=G_i$ to find the amplitudes $A_k$ and $f_\mathrm{r,0}$.
	\item Adjusting the phase using $\varphi_{i+1}=\alpha(\varphi_\mathrm{ext} - \frac{2\pi}{\Phi_0}L_\mathrm{loop}I_\mathrm{J}(\varphi_i)) + (1-\alpha)\varphi_i$, where we introduce the solution relaxation parameter $0<\alpha<1$ when updating the phase to ensure stable convergence (We checked that the value of $\alpha$ does not affect the result).
	\item We update the microwave shunt conductance $G_{i+1} = \mathrm{Re}[(Y_\mathrm{m}^{-1} - j{\omega}L_\mathrm{loop})^{-1}]$, using $Y_\mathrm{m}$ calculated via Eqs.~\ref{seq:f_res} and \ref{seq:q_i} for $f_\mathrm{r,0}$ obtained in step 1.
\end{enumerate}
We use $\alpha=0.3$ and do a fixed number of iterations $n_\mathrm{iter}=40$. This proves enough to ensure the convergence of $A_k$ and $f_\mathrm{r,0}$. Prior to the iterative fit, the magnetic field periodicity and offset in $f_\mathrm{r}$ and $Q_\mathrm{i}$ have been determined using cross-correlation and mean squares minimization of $f_\mathrm{r}$ and $Q_\mathrm{i}$.

Figure~\ref{sfig:admittance_d1} illustrates the results of the fit procedure for device D1. Here, self-consistent extraction is applied for all gate voltages for $T>20\,\unit{\milli\kelvin}$, as well as for $V_\mathrm{g}=-2.5\,\unit{\volt}$ and $V_\mathrm{g}=-1.5\,\unit{\volt}$ at $T=20\,\unit{\milli\kelvin}$, where no hysteresis is observed.

\subsection*{Analysis of the RF SQUID hysteresis}

The self-consistent procedure described above fails to converge when RF SQUID hysteresis occurs, as indicated by the condition $L_\mathrm{J}^{-1}(\varphi) < -L_\mathrm{loop}^{-1}$. This failure arises because the dependence of $\varphi$ on $\varphi_\mathrm{ext}$ is no longer continuous.

To extract the CPR in this hysteretic regime, we employ an alternative method:
\begin{enumerate}
	\item We calculate $Y_\mathrm{m}(\varphi_\mathrm{ext})$ using Eqs.~\ref{seq:f_res} and \ref{seq:q_i} relying on $f_\mathrm{r,0}$ extrapolated from measurements conducted in the absence of RF SQUID hysteresis, where the self-consistent extraction procedure was applicable.
	\item The Josephson junction admittance as a function of external phase bias $Y(\varphi_\mathrm{ext})$ is then computed using $Y(\varphi_\mathrm{ext})=[(Y_\mathrm{m}(\varphi_\mathrm{ext}))^{-1} - j{\omega}L_\mathrm{loop}]^{-1}$. The relationship between $Y$ and $Y_\mathrm{m}$ is independent of flux screening, as described by Eq.~\ref{seq:chi_screening}.
	\item We fit $\mathrm{Im}[Y(\varphi_\mathrm{ext})]$ using Eqs.~\ref{seq:cpr_ansatz} and \ref{seq:josephson_inductance} to compute $\mathrm{Im}[Y(\varphi)]$ and Eq.~\ref{seq:flux_screening} to determine $\varphi_\mathrm{ext}(\varphi)$, explicitly accounting for the phase jumps. The phase is then interpolated to produce $\mathrm{Im}[Y(\varphi_\mathrm{ext})]$ for the fit.
\end{enumerate}

Due to the computational intensity of this calculation, we restrict the analysis to the five lowest harmonics in Eq.~\ref{seq:cpr_ansatz}.

Since the hysteresis condition $L_\mathrm{J}^{-1}(\varphi) < -L_\mathrm{loop}^{-1}$ directly affects the magnitude of the measured signal, achieving a satisfactory fit requires accurate tuning of the effective mutual inductance $M$ used in Eqs.~\ref{seq:f_res} and \ref{seq:q_i}. We manually adjust $M$ to optimize the fit, thereby calibrating the measurement.

The results of this procedure are presented in Fig.~\ref{sfig:admittance_d1} for $T=20\,\unit{\milli\kelvin}$ and $V_\mathrm{g}=-1.0\,\unit{\volt},\,-0.5\,\unit{\volt}$ and $-0.2\,\unit{\volt}$. In Fig.~\ref{sfig:admittance_d1}(c-e), the solid lines represent the experimentally accessible $\mathrm{Im}[Y]$ where the phase $\varphi$, while the dashed line are extracted from a fit assuming a CPR composed of the five lowest sine harmonics. The corresponding CPRs is shown in Fig.~\ref{sfig:admittance_d1}(m-o). Additionally, $\mathrm{Re}[Y]$ is shown in Fig.~\ref{sfig:admittance_d1}(h-j).

Using the RF SQUID hysteresis and the estimated loop inductance we obtain the effective mutual inductance between SQUID loop and the resonator. For D1, we obtain the value $M\approx9.45\,\unit{\pico\henry}$, while for D2 $M\approx13.1\,\unit{\pico\henry}$. For device D3 no hysteresis is observed due to the smaller loop inductance, and we estimate $M=3.5\,\unit{\pico\henry}$ based on the supercurrent magnitude of device D4. This device was fabricated alongside D3 from the same wafer and has identical geometry.

For flip-chip bonded devices, $M$ can also be estimated based on the distances measured from the images taken during alignment. The specifics of the inductive coupling of the loop to the resonator must be taken into account, which we discuss in Supplemental Material, Note 8.

\newpage
\section{Inductance of the RF SQUID loop and mutual inductance}

\subsection*{Inductance of the RF SQUID loop}

We calculate the geometric inductance of our rectangular loops, $L_\mathrm{geom}$, using Sonnet. We obtain $L_\mathrm{geom}\approx171\,\unit{\pico\henry}$ for the $70\,\unit{\micro\meter}\times47\,\unit{\micro\meter}$ loop, and $L_\mathrm{geom}\approx72.2\,\unit{\pico\henry}$ for the $40\,\unit{\micro\meter}\times20\,\unit{\micro\meter}$ loop.

The kinetic inductance contribution is relevant in our case, so that $L_\mathrm{loop}= L_\mathrm{geom} + N_{\square}L_\square$, where $L_\square$ is the kinetic inductance per square and $N_{\square}$ is the number of squares in a loop trace. We calculated $N_\square\approx160$ for the larger loop, and $N_\square\approx84$ for the smaller loop.

The presence of normal metal layers in the electrode stack reduces the superconducting gap in the structure and increases kinetic inductance. The metal stack of our superconducting electrodes consists of a superconductor with a larger gap, $\mathrm{Al}$, a superconductor with a small gap, $\mathrm{Ti}$, and the normal metal capping layer, $\mathrm{Au}$. We thus expect a larger $L_\square$ than that for pure $\mathrm{Al}$. 

We observe the effect of the electrode stack in the DC measurements on D4 (Appendix A), where we extract a superconducting gap $\Delta_0\approx136\,\unit{\micro\electronvolt}$, which is suppressed compared to the gap value in pure Al, $\Delta_\mathrm{Al}\approx180\,\unit{\micro\electronvolt}$. As the ratio $\Delta_0/\Delta_\mathrm{Al}$ in our material is comparable to the results in Ref.\,\cite{S_Hu2020} for a bilayer Al/Au stack, we assume a similar value for the kinetic inductance per square, $L_\square\approx0.5\,\unit{\nano\henry}$. We disregard the effect of interfaces. These approximations ultimately limit the accuracy of $I_\mathrm J$ and $G$. With $L_\square=0.5\,\unit{\pico\henry}$, we obtain $L_\mathrm{loop}\approx251\,\unit{\pico\henry}$ for D1 and D2, and $L_\mathrm{loop}\approx114\,\unit{\pico\henry}$ for D3.

\subsection*{Mutual inductance between the loop and the resonator}

\begin{figure*}[h]
	\centering
	\includegraphics[scale=0.6]{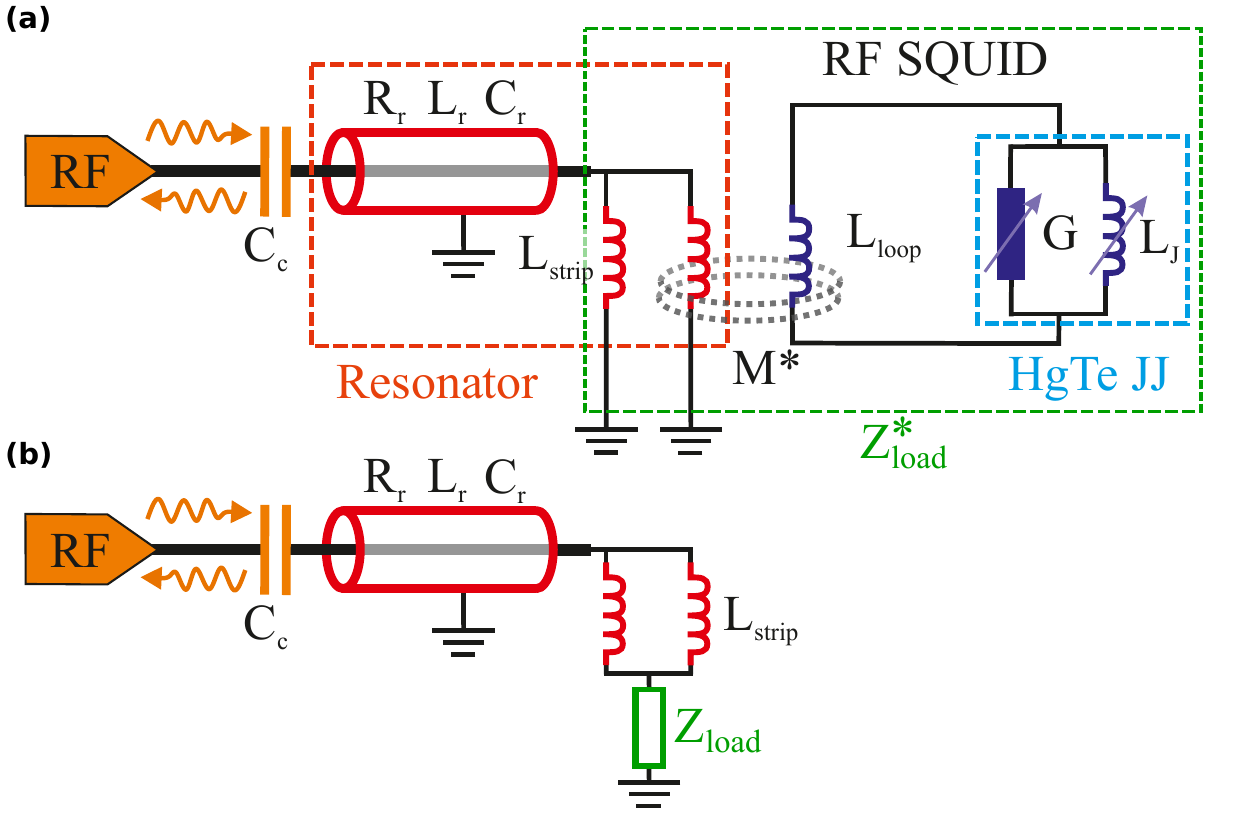}
	\caption{\textbf{Equivalent circuit of the RF SQUID inductively coupled to the resonator.} \textbf{(a)} Mechanism of inductive coupling with the RF SQUID loop. The junction (blue) is modeled by a variable Josephson inductance, $L_\mathrm{J}$, and a phase dependent shunt conductance, $G$. The resonator (red) is formed by a transmission line with distributed parameters $R_\mathrm{r}$, $L_\mathrm{r}$, $C_\mathrm{r}$, shorted at the far end with two inductive strips. RF SQUID loop is coupled to one of those strips with the geometric mutual inductance $M^*$. The other end of the resonator is connected to a transmission line through a coupling capacitor $C_\mathrm{c}$. The elements forming variable load impedance ($Z_\mathrm{load}^*$) at the shorted resonator end are inside the green box. \textbf{(b)} Equivalent circuit, used for the analysis of microwave response. The effective variable load impedance ($Z_\mathrm{load}$) is inserted between both inductive strips and the ground. Transition to the equivalent circuit (b) results in twofold reduction of the effective mutual inductance $M=M^*/2$
	}
	\label{sfig:equivalent_circuit}
\end{figure*}

In this section, we discuss the coupling of the RF SQUID loop and the resonator and obtain a value for the effective mutual inductance $M$. Eqs.~\ref{seq:f_res} and~\ref{seq:q_i} are derived for a configuration, where the RF SQUID loop is inductively coupled to a single strip at the grounded end of the resonator, such that the effective variable load impedance $Z_\mathrm{load}=Y_\mathrm{m}{\omega}^2M^2$ is inserted between the resonator and the ground~\cite{S_Haller2022}.

The design of our resonator is slightly different [Fig.~\ref{sfig:flip_chip_devices}(b)] as it is grounded via two identical thin strips connected to a triangular pad. The RF SQUID loop is inductively coupled to only one of these strips, so that effective circuit looks closer to Fig.~\ref{sfig:equivalent_circuit}. The variable load at the resonator end then includes two parallel strips with inductance $L_\mathrm{strip}$ each, one of which has the impedance $Y_\mathrm{m}{\omega}^2{M^*}^2$ connected in series. Here $M^*$ is the geometric mutual inductance that is calculated based on the geometry of loop and strip from basic principles.

The variable load impedance at the resonator end is calculated to be: 
\begin{equation}
	{Z_\mathrm{load}^*}^{-1} = (j{\omega}L_\mathrm{strip})^{-1} + (j{\omega}L_\mathrm{strip} + Y_\mathrm{m}{\omega}^2{M^*}^2)^{-1}.
	\label{eq:variable_load}
\end{equation}
Assuming only a geometrical inductance contribution, we find $L_\mathrm{strip}>188\,\unit{\pico\henry}$ (However, $L_\mathrm{strip}$ can also include a notable contribution from the kinetic inductance).

Considering $Y_\mathrm{m}\sim0.1\,\unit{\ohm^{-1}}$, $M\sim20\,\unit{\pico\henry}$, this results in a condition $|j{\omega}L_\mathrm{strip}|\gg |Y_\mathrm{m}{\omega}^2{M^*}^2|$ being satisfied for all measurement frequencies. This allows to simplify Eq.~\ref{eq:variable_load} in two steps:

\begin{equation}
	\begin{split}
		& {Z_\mathrm{load}^*}^{-1} \approx 2 (j{\omega}L_\mathrm{strip})^{-1} - Y_\mathrm{m}{\omega}^2{M^*}^2 (j{\omega}L_\mathrm{strip})^{-2} \\
		& {Z_\mathrm{load}^*} \approx (j{\omega}L_\mathrm{strip})/2 + Y_\mathrm{m}{\omega}^2{M^*}^2 / 4
	\end{split}
	\label{eq:variable_load_simplified}
\end{equation}

The part of the load impedance pertaining to strip [$(j{\omega}L_\mathrm{strip})/2$] is independent of the coupled SQUID loop and thus may be absorbed into the resonator inductance. The remaining $Y_\mathrm{m}{\omega}^2{M^*}^2 / 4$ yields $M=M^*/2$ when compared with $Z_\mathrm{load}=Y_\mathrm{m}{\omega}^2M^2$. The effective mutual inductance $M$ is then twice smaller than the geometric $M^*$ as we couple to only one of strips grounding the resonator.
\newpage
\section{Comparison between microwave loss at phase 0 and I$_\mathrm{c}$}

\begin{figure*}[h]
	\centering
	\includegraphics[scale=1]{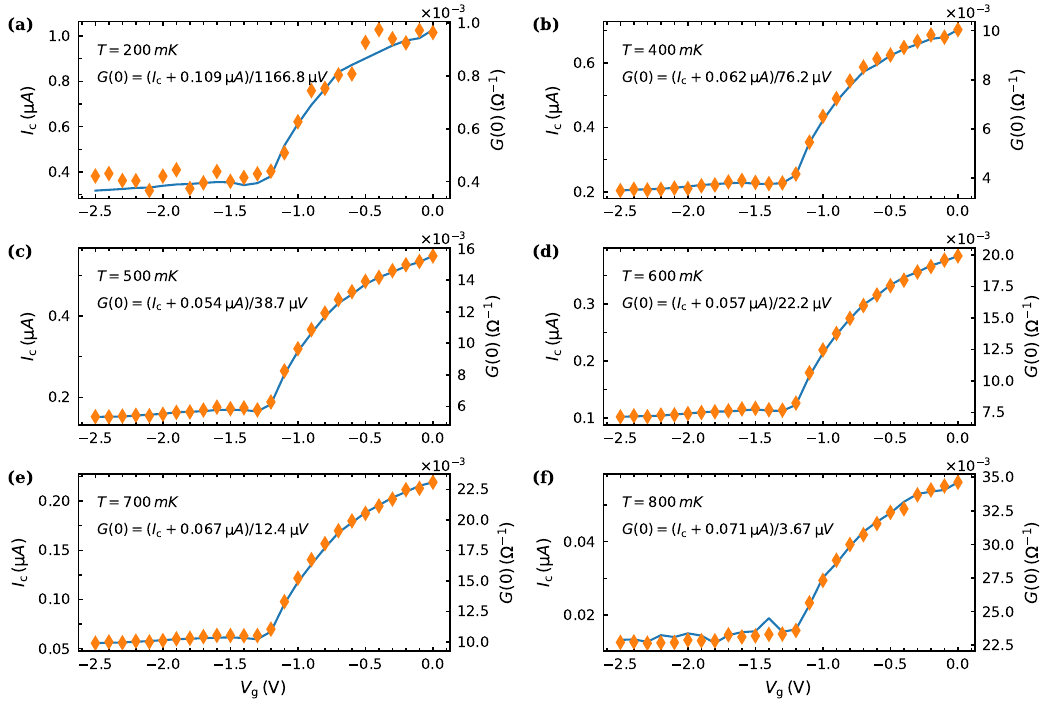}
	\caption{
		\textbf{Gate voltage dependencies of $I_\mathrm{c}$ and $G(0)$ at different temperatures.}
		In all panels (a-f) the left axis corresponds to $I_\mathrm{c}$ magnitude (blue lines), whereas the right axis corresponds to $G(0)$ (orange symbols). The equation in the panel shows the obtained relation between $G(0)$ and $I_\mathrm{c}$
	}
	\label{sfig:compare_i_c_g}
\end{figure*}

At all temperatures where $G(0)$ is pronounced, we observe exact matching between the gate voltage dependencies of $G(0)$ and critical current $I_\mathrm{c}$. We demonstrate this in Fig.~\ref{sfig:compare_i_c_g} for six selected temperatures. In each case, we establish a relation between $I_\mathrm{c}$ and $G(0)$ in a form $G=(I_\mathrm{c}+a)/b$, with parameter $a\sim50\,\unit{\nano\ampere}$ having weak temperature dependence and strongly $T$-dependent $b$.

\section{Experimental Results for the second 500\,\lowercase{nm} Device D2}

\begin{figure*}[h]
	\centering
	\includegraphics[scale=1.0]{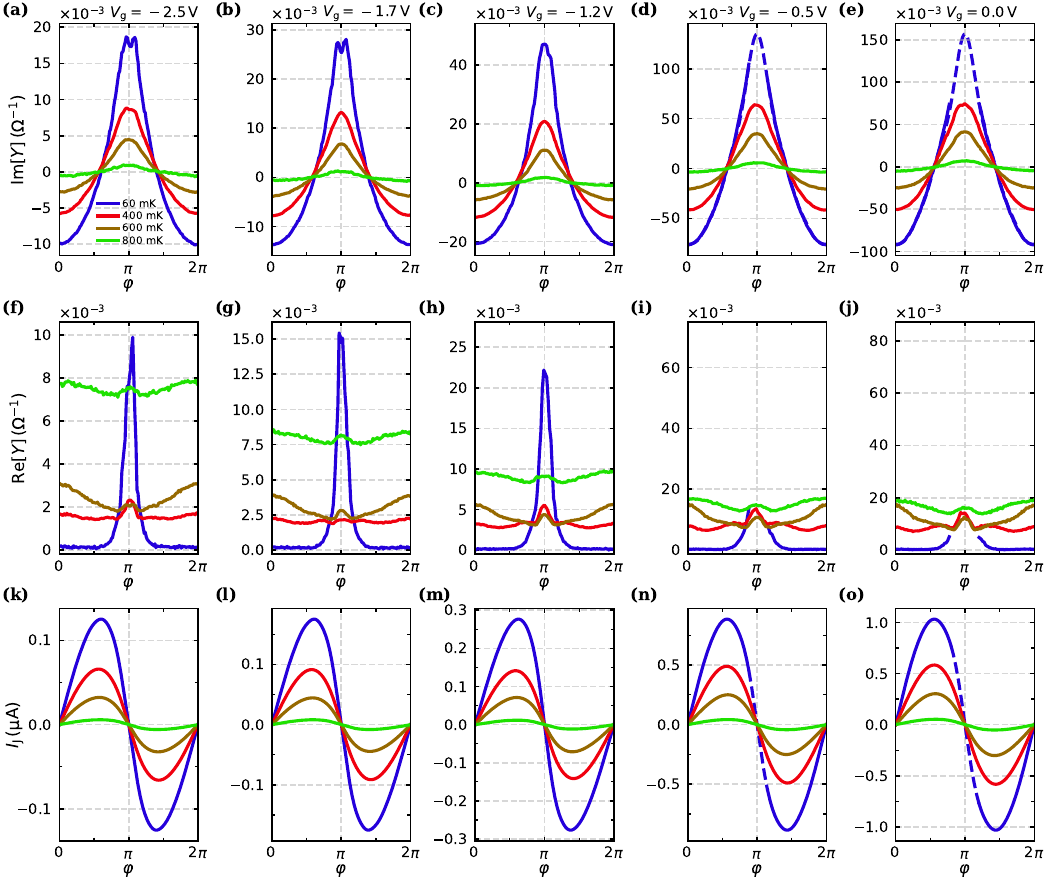}
	\caption{\textbf{Components of the microwave admittance $Y$ for device D2.} \textbf{(a-e)} Imaginary component $\mathrm{Im}[Y]$ as a function of JJ phase $\varphi$, primarily related to the Josephson inductance $L_\mathrm{J}^{-1} = \frac{2\pi}{\Phi_0}\frac{\partial I_\mathrm{J}}{\partial \varphi}$ for several temperatures [legend in panel (a)]. The dashed lines in some $T=60\,\unit{\milli\kelvin}$ traces mark the region where the admittance is inaccessible and is obtained from a fit assuming a CPR composed of the five lowest sine harmonics. \textbf{(f-j)} Real part of microwave response, $\mathrm{Re}[Y]$ labeled throughout the manuscript as the microwave shunt conductance $G$. \textbf{(k-o)} Current-phase relation (CPR), extracted from $\mathrm{Im}[Y]$ in (a-e) using the self-consistent extraction procedure.
	}
	\label{sfig:admittance_d2}
\end{figure*}

\begin{figure*}[h]
	\centering
	\includegraphics[scale=1]{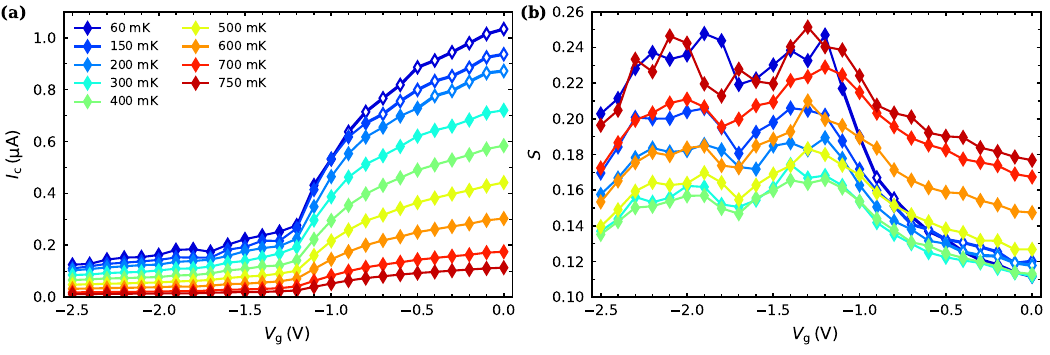}
	\caption{
		\textbf{Gate dependence of critical current and skewness for device D2.}
		\textbf{(a)} Critical current $I_\mathrm{c}$ as a function of gate voltage $V_\mathrm{g}$, extracted from microwave measurements at different temperatures (see legend). Empty symbols represent points where SQUID hysteresis is present.
		\textbf{(b)} Skewness of the current-phase relation $\beta_\mathrm{s}$ measured alongside the data in (a). Empty symbols on the right are less precise, since, again, part of the current-phase relation can not be measured there (see main text).
	}
	\label{sfig:d2_ic_beta_vs_vg}
\end{figure*}

\begin{figure*}[h]
	\centering
	\includegraphics[scale=1]{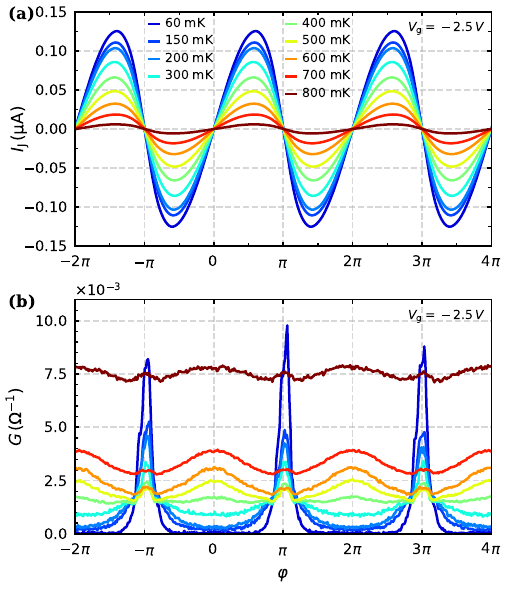}
	\includegraphics[scale=1]{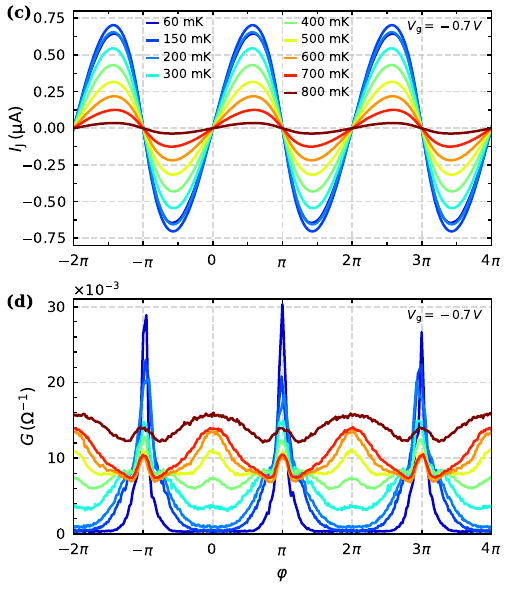}
	\caption{
		\textbf{CPR and phase-dependent loss for device D2.}
		\textbf{(a,b)} Supercurrent $I_\mathrm{J}$ (a) and microwave loss at gate voltage $V_\mathrm{g}=-2.5\,\unit{\volt}$ as a function of phase bias $\varphi$ at different temperatures (see legend).
		\textbf{(c,d)} Same as (a,b), but for $V_\mathrm{g}=-0.7\,\unit{\volt}$.
	}
	\label{sfig:d2_ij_phase}
\end{figure*}

Here we present the measurement results for a second $500\,\unit{\nano\meter}$ long device, D2. Device D2 has a lower critical current at $V_\mathrm{g}=0$ and can be gated to smaller currents than D1 [Fig.\,\ref{sfig:d2_ic_beta_vs_vg}(a)]. This is likely the result of a lower carrier density being introduced during fabrication. At the same time, the observations reported in main text are valid for D2 as well. We observe notably forward-skewed CPR [Fig.\,\ref{sfig:d2_ij_phase}(a,c)], with skewness $\beta_\mathrm{s}$ having small change $\approx30\%$ with $V_\mathrm{g}$ and $T$ [Fig.\,\ref{sfig:d2_ic_beta_vs_vg}(b)]. In this measurement, the lowest available temperature is $T=60\,\unit{\milli\kelvin}$. Similarly to D1, at certain $V_\mathrm{g}$ and $T$, we observe RF SQUID hysteresis (empty symbols).

At all gate voltages, the phase dependent microwave loss behaves similarly to the sample discussed in the main text, i.e., at low temperatures we find a loss peak at $\varphi=\pi$ and a near zero shunt conductance $G$ at $\varphi=0$. At higher temperatures, the size of the peak diminishes, whereas the loss at $\varphi=0$ increases [Fig.\,\ref{sfig:d2_ij_phase}(b,d)].

\clearpage
\section{Experimental Results for the 200\,\lowercase{nm} Device D3}

\begin{figure*}[h]
	\centering
	\includegraphics[scale=1]{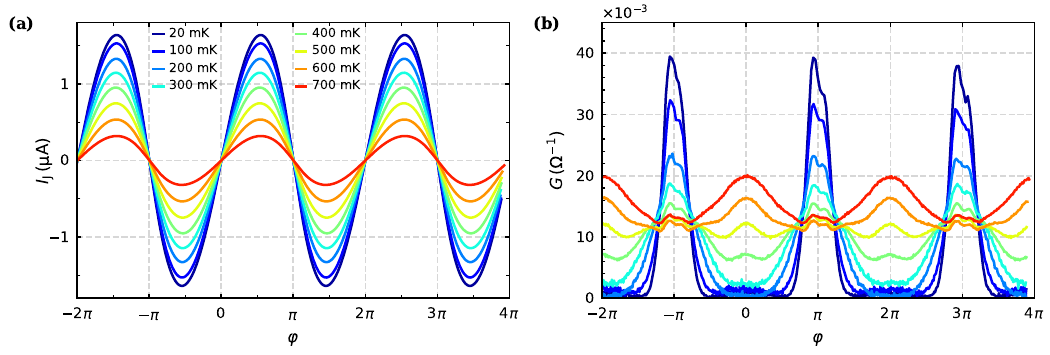}
	\caption{\textbf{Measured microwave admittance of device D3 at $V_\mathrm{g}=0\,\unit{\volt}$.} \textbf{(a)} Current phase relation at different temperatures. \textbf{(b)} Phase-dependent microwave loss at different temperatures.}
	\label{sfig:microwave_suscept}
\end{figure*}

In Fig.~\ref{sfig:microwave_suscept}, we show the current-phase relation and microwave loss of device D3 at $V_\mathrm{g}=0\,V$ for different temperatures. We observe the dissipation peak at even phases $\varphi_\mathrm{even}$ appearing for temperatures below $T\approx400\,\unit{\milli\kelvin}$.

\end{document}